\documentclass{elsart}

\usepackage{psfig}
\usepackage{graphicx}
\usepackage{amsmath}
\usepackage{amssymb}
\usepackage{rotating}
\usepackage{pstricks}
\usepackage{pst-plot}
\usepackage{pst-node}
\usepackage{url} 

\journal{Physica A}

\begin{document}

\begin{frontmatter}
\title{On the Equilibrium Fluctuations of an Isolated System}

\author{Kieran Kelly},
\ead{kkelly@probabilitydynamics.com}
\author{Przemys{\l}aw Repetowicz}, \&
\ead{prepetowicz@probabilitydynamics.com}
\author{Seosamh macR{\'e}amoinn}
\ead{smacreamoinn@probabilitydynamics.com} 

\address{Probability Dynamics, IFSC House, Custom House Quay, Dublin 1, Ireland.}
 
\date{\today}

\begin{abstract}
Traditionally, it is understood that fluctuations in the equilibrium distribution are not evident in thermodynamic systems of large $N$ (the number of particles in the system) \cite{Huang1}. In this paper we examine the validity of this perception by investigating whether such fluctuations can in reality depend on temperature. 

Firstly, we describe fluctuations in the occupation numbers of the energy levels
for an isolated system, using previously unknown identities that 
we have derived for the purpose, which allow us to calculate the moments of the occupation numbers. Then we compute analytically the probability distribution of these fluctuations. We show that, for every system of fixed and finite $N$, fluctuations about the equilibrium distribution do, in fact, depend on the temperature. Indeed, at higher temperatures the fluctuations can be so large that the system does not fully converge on the Maxwell-Boltzmann distribution but actually fluctuates around it. We term this state, where not one macrostate but a region of macrostates closely fit the underlying distribution, a ``{\it fluctuating equilibrium}''. Finally, we speculate on how this finding is applicable to networks, financial markets, and other thermodynamic-like systems.
\end{abstract}

\end{frontmatter}

\section{Introduction}

The relaxation of a classical statistical system,  
that is isolated from its surroundings, towards the equilibrium Maxwell-Boltzmann distribution in the thermodynamic limit is well understood. See Huang~\cite{Huang1} for the derivation of that distribution by means of Lagrange's method of undetermined multipliers and Friedman et al.~\cite{Friedman} and references therein for more pedagogical derivations aimed at an audience that is not versed in multivariate differential calculus.
As fluctuations in this distribution effectively all but disappear in the thermodynamic limit, this distribution has proven sufficient for describing many thermodynamic systems for which this limit applies.

In this paper, we consider an isolated system with a \textit{large but finite} number of particles, $N$. Much work has been done on finite systems, away from the thermodynamic limit. For example,  Hill~\cite{Hill} has presented a wider scope for thermodynamics, that of describing small systems, while Lebowitz et al.~\cite{Lebowitz:s} have analysed finite size effects using Taylor series expansions. More recent work has included topics such as the nonequivalence of thermodynamic ensembles~\cite{Challa, Gulminelli}, and phase transitions~\cite{Gulminelli,Kastner,Dunkel,Gross}. 

However, in our work, we set out to derive the distribution of particles among energy levels/states of an isolated system with a large but finite number of particles, $N$, and the finite-$N$ corrections to the distribution using a simple model.
Kelly~\cite{Kelly} has recently investigated this problem numerically, and macR\'{e}amoinn~\cite{Seosamh} has computed analytically the variance of the distribution (the amount of fluctuations). 
We now find closed form expressions for all the moments of the distribution, along with the multi-variate distribution of the occupation numbers. We show that at higher temperatures the amount of fluctuations exceeds the mean occupation number and thus no stable equilibrium distribution is attained. 
We describe this phenomenon as a ``\textit{fluctuating equilibrium}''
and we hypothesise that it could be useful to describe certain thermodynamic or thermodynamic-like systems, for example, in the global financial markets.

\section{Theoretical Analysis} 

Our approach follows closely the permutational argument of Boltzmann~\cite{Boltzmann} and is similar to methods illustrated (by examples) in textbooks~\cite{Blatt,Eisberg,Tipler}. However, we want more than just the most probable distribution, and we do not take the limit of continuity in our calculations.

We consider an ideal gas of $N$ particles in a volume $V$ ($\frac{N}{V} << 1$) where the mutual energy of interaction between the molecules is negligibly small. We assume that the collisions with walls of container and between molecules are perfectly elastic, and that the walls are rigid. For an ideal gas, the energy of the system, $E$, is given by:
\begin{equation}
\label{eq:total_energy_ig} \mbox{$E = \frac{3}{2}N k_{B} T$}
\end{equation}
where $k_B$ is Boltzmann's constant, and $T$ the absolute temperature.
\footnote{Of course, in general, temperature is normally defined as the partial derivative of the 
internal energy with respect to entropy with $N$ and $V$ held constant.}. 
We assume the system is isolated from its 
surroundings. Following Boltzmann himself~\cite{Boltzmann}, we 
assume the possible energies of the particles are given by equidistant, non-degenerate energy levels (with energy level spacing $\epsilon$), with the lowest 
energy level set to zero for simplicity.  Then the energy of each particle can be 
written as a multiple of the energy level spacing, $j \epsilon$, where $j \in \lbrace 
0, \ldots M\rbrace$ \footnote{Unlike Boltzmann, we only consider the case where the maximum number of energy units a particle can have is equal to the total number of units in the system, $M$. Clearly, for large $N$, the likelihood of a
particle having energies of high $j$ ($j \sim M$) becomes very small.}. Thus we have:
\begin{equation}
\label{eq:total_energy_els} \mbox{$E = M\epsilon$} 
\end{equation}
We now use a permutational argument to describe, in a quantitative fashion, both the most likely distribution of the particles among the energy states and the fluctuations around this distribution.

\subsection{The Micro- \& Macrostates}

We assume that the particles are distinguishable, as Boltzmann did.
\footnote{Distinguishability would be a good approximation for thermodynamic systems away from the quantum regime of low temperature or high density. 
The resulting (Maxwell Boltzmann) statistics gives a non-extensive entropy but this does not concern us for the purposes of this paper. 
Of course, there has been plenty of discussion on entropy being extensive~\cite{Jaynes,Addison,Sheehan}. 
Also, we believe that this model may be suitable for economic systems where distinguishability is valid.}
We describe a microstate as a certain distribution of particles among the energy levels where we distinguish between the different particles (e.g. by assigning labels to the particles) - Boltzmann called them ``complexions''. 
We define a macrostate as a coarse-graining of this, where we are only interested in the number of particles occupying each energy level.

We denote by $n_j$, the occupation number of the $j^{th}$ energy level, and specify the distribution of particles among the energy levels (the macrostates) by \mbox{$\vec{n} := \left(n_j\right)_{j=0}^M$}. The total number of particles and the total energy are conserved i.e.  
\begin{equation}
N = \sum\limits_{j=1}^M n_j
\quad\mbox{and}\quad M = \sum\limits_{j=0}^M j n_j
\label{eq:ConservationLaws}
\end{equation}
In this paper we are interested in the probability of macrostates. By assuming each microstate is equally likely, we can assign (unnormalized) probabilities to the macrostates by the number of microstates corresponding to one macrostate. 
This is clearly given by the following multinomial factor:
\begin{equation}
\frac{N!}{\prod\limits_{j=0}^M n_j!}
\label{eq:Multiplicity}
\end{equation}
since it is the number of ways we can place $n_j$ particles on the $j^{th}$ energy level for $j=0,\dots,M$.

The total number of microstates of the system $\left(n(\mbox{micro})\right)$ reads:
\begin{eqnarray}
n(\mbox{micro}) &=& \sum\limits_{\sum\limits_{j=0}^M n_j=N} 
\frac{N!}{\prod\limits_{j=0}^M n_j!} \delta_{M,\sum\limits_{j=0}^M j n_j} \\
& =&
\int\limits_0^{2 \pi} \frac{d\phi}{2\pi} e^{\imath \phi M} 
\left(\sum\limits_{j=0}^M e^{-\imath j \phi}\right)^N 
\label{eq:Microstates1} \\
&=&\oint \frac{dz}{2\pi \imath z} z^{-M} \left(\frac{1- z^{M+1}}{1 - z}\right)^N 
\label{eq:Microstates1A}\\
&=& \left.\frac{1}{M!} \frac{d^M}{d z^M} \left(\frac{1- z^{M+1}}{1 - z}\right)^N\right|_{z=0} \label{eq:Microstates1B} \\
\Rightarrow n(\mbox{micro}) &=& C^{M+N-1}_{N-1} 
 = \frac{(M+N-1)!}{M! (N-1)!}
\label{eq:Microstates2} 
\end{eqnarray}
In (\ref{eq:Microstates1}) we inserted the integral representation of the delta function, and in (\ref{eq:Microstates1A}) and (\ref{eq:Microstates1B}) 
we substituted for $e^{-\imath \phi}$ and we computed the resulting integral using the Cauchy theorem.
The final equality can be proven by the generalized Leibnitz rule and by induction in $N$.
\footnote{We have also derived a closed form expression for the total number of macrostates. 
However since we focus in this work on fluctuations around the equilibrium we will present that result elsewhere.}

\subsection{Fluctuations Around Equilibrium}

The most likely  macrostate is obtained by maximizing the multiplicity of macrostates 
(\ref{eq:Multiplicity})
subject to conditions  (\ref{eq:ConservationLaws}). In the thermodynamic limit ($N\rightarrow \infty$)
this yields the Maxwell-Boltzmann exponential distribution of energies 
with the decay constant proportional to the inverse of the absolute temperature $T$, as shown in \cite{Huang1}, for example. 
For an ideal gas the absolute temperature is proportional
 to the mean kinetic energy of the particles which, in our model, is proportional to the ratio
 $\frac{M}{N}$. This can be seen clearly from equations~(\ref{eq:total_energy_ig})
and~(\ref{eq:total_energy_els}). In fact, we have:
\begin{equation}
\label{eq:temp_def} \mbox{$\frac{M}{N} = \frac{3 k_B}{2 \epsilon} T$}
\end{equation}
In what follows, we use ``temperature'' and ``$T$'' for this ratio $\frac{M}{N}$. 
Of course, this ratio is the ``specific energy'', simply the average energy per particle in multiples of the energy levels spacing $\epsilon$, which is a well defined quantity in itself, no matter how we define temperature.

While the most likely distribution is well known, little is known about the fluctuations around this distribution.
This raises interesting questions. The traditional understanding of statistical mechanics suggests to us that an isolated thermodynamic system will always move 
to the Maxwell-Distribution as a result of the Law of Large Numbers (LLN). 
We wish to investigate the effect of the temperature (the ratio $\frac{M}{N}$) on this convergence. 
If the energy content of the system is ``very large'' (see later for a comment on what this might mean), such that the temperature is high, 
maybe the best the system can achieve is only an equilibrium \textit{region of states}; ``a fluctuating equilibrium''.
Indeed, at very high temperatures, the fluctuations could be so dominant that disorder and chaos would prevail at the macroscopic level. 
We believe such behaviour may be apparent in thermodynamic-like systems where $N$ is much smaller than Avagadro's number.

In order to investigate this possibility we must first compute all the moments of the occupation numbers of the energy levels. We have:
\begin{eqnarray}
\lefteqn{C^{M+N-1}_M \left<n_j^m\right> := \sum\limits_{\sum\limits_{p=0}^M n_p=N} n_j^m \frac{N!}{\prod\limits_{p=0}^M n_p!}  \delta_{M,\sum\limits_{p=0}^M p n_p}}
\label{eq:Moments1} \\
&&= 
\left.\frac{d^m}{d\log(x)^m} \int\limits_0^{2\pi} \frac{d\phi}{2\pi} e^{\imath M \phi}
\sum\limits_{\sum\limits_{p=0}^M n_p=N}\frac{N!}{\prod\limits_{p=0}^M n_p!} \prod\limits_{p=1}^M (e^{-\imath \phi})^{p n_p} (\delta_{p,j} (x^{n_j}-1) + 1))
\right|_{x=1}
\label{eq:Moments2} \\
&&=
\left.\frac{d^m}{d\log(x)^m} \int\limits_0^{2\pi} \frac{d\phi}{2\pi} e^{\imath M \phi}
\left(\frac{1-(e^{-\imath\phi})^{M+1}}{1- e^{-\imath\phi}} + (x-1)e^{-\imath \phi j}\right)^N
\right|_{x=1}
\label{eq:Moments3} \\
&&=
\left.\left.\frac{d^m}{d\log(x)^m} \frac{1}{M!} \frac{d^M}{dz^M}
\left(\frac{1-z^{M+1}}{1-z} + z^j(x-1)\right)^N
\right|_{z=0}
\right|_{x=1}
\label{eq:Moments4} \\
&&=
\left.\frac{d^m}{d\log(x)^m} 
\sum\limits_{q=0}^{N-1} 1_{q j \le M} (x-1)^q C^N_q C^{M-q j + N-1 -q}_{N-1-q}
+ \delta_{N j,M} (x-1)^N
\right|_{x=1}
\label{eq:Moments5} \\
&&=
\sum\limits_{q=1}^{(N-1)\wedge m} 1_{q j \le M} a_q^{(m)} q! C^N_q C^{M-q j + N-1 -q}_{N-1-q}
+ \delta_{N j,M} N! a_N^{(m)}
\label{eq:Moments6}
\end{eqnarray}
In (\ref{eq:Moments2}) we introduced an integral representation of the delta function and we used the identity:
\begin{equation}
n^m = \left.\frac{d^m}{d\log(x)^m} x^n\right|_{x=1}
\end{equation}  
for $m$ a non-negative integer. In (\ref{eq:Moments3}) we computed the sum using the multinomial expansion formula
and in (\ref{eq:Moments4}) we substituted for $z=e^{-\imath \phi}$ and computed the resulting complex
integral using the Cauchy theorem. 
In (\ref{eq:Moments5}) we computed the derivative using identity (\ref{eq:PowerofSum})
and in (\ref{eq:Moments6}) we used the identity (\ref{eq:DiffExp}). The coefficients $a_q^{(m)}$ are defined in 
(\ref{eq:Coefficients})-(\ref{eq:Coefficients2}).

From (\ref{eq:Moments6}) we see that for $N\rightarrow \infty$ the moments of the occupation densities of the energy levels, $x_j := n_j/N$, read:
\begin{equation}
\left<x_j^m\right> = 1_{m j\le M}  \frac{C^{M-m j + N-1 -m}_{N-1-m}}{C^{M+N-1}_{N-1}} + O(\frac{1}{N})
\label{eq:MthMomentLimit}
\end{equation} 
and are all finite, once the temperature $T = M/N$, is finite.

Now imagine that $N \rightarrow \infty$ but $T$ is kept finite. Then from (\ref{eq:MthMomentLimit}) we have:
\begin{eqnarray}
\left<x_j^m\right> &=& 1_{m j\le M}\frac{\left(\prod\limits_{p=1}^m (N-p)\right)\left(\prod\limits_{p=0}^{m j-1} (M-p)\right)}{\prod\limits_{p=1}^{m j + m} (M+N-p)}\nonumber  \\
& \mathop{=}_{N \to \infty}&
\left(\frac{N}{M+N}\right)^m \left(\frac{M}{M+N}\right)^{m j}
\nonumber \\
&=&
\frac{1}{(1+T)^m} e^{-j m \log(1+T^{-1})} \nonumber \\
&=&
\left(\frac{T^j}{(T+1)^{j+1}}\right)^m
\label{eq:MthMomentLimitResult}
\end{eqnarray} 
and we conclude that the moments of the distribution depend on 
both  the energy of the level $j$ and on the temperature of the system.
In the case $m=1$ we retrieve results from \cite{Friedman} (equation (8), page 119) and from \cite{Kelly} (equation (6), page 10).
In the case $m=2$ the result fits in with \cite{Seosamh} (equation (33), page 9).

Thus all the moments, and in turn also the distributions of the occupation densities, depend
on  the temperature $T$ only, and not on either $N$ or $M$ alone.
In order to check that the formula is correct we use MATHEMATICA to show that
the total number of particles is conserved:
\begin{equation}
\sum\limits_{j=0}^M \left<x_j\right> = 
\sum\limits_{j=0}^M \frac{C^{M-j+N-2}_{N-2}}{C^{M+N-1}_{N-1}}
=1
\end{equation}
For the limit $N \to \infty$ with fixed $T$ (which means that $M \to \infty$ as $N = TM$) we get:
\begin{eqnarray}
\sum\limits_{j=0}^M \left<x_j\right> &=& \frac{1}{(1+T)} \sum\limits_{j=0}^M  e^{-j  \log(1+T^{-1})} 
\nonumber \\
&=&
\frac{1}{(1+T)} \frac{1 - (\frac{T}{1+T})^{M+1}}{1 - \frac{T}{1+T}} \nonumber\\
&=& 1 - (\frac{T}{1+T})^{M+1} \mathop{\simeq}_{M\to \infty} 1
\label{eq:NormalizationLimit}
\end{eqnarray} 
as expected. 

\begin{figure}[tbh]
\hbox{\psfig{figure=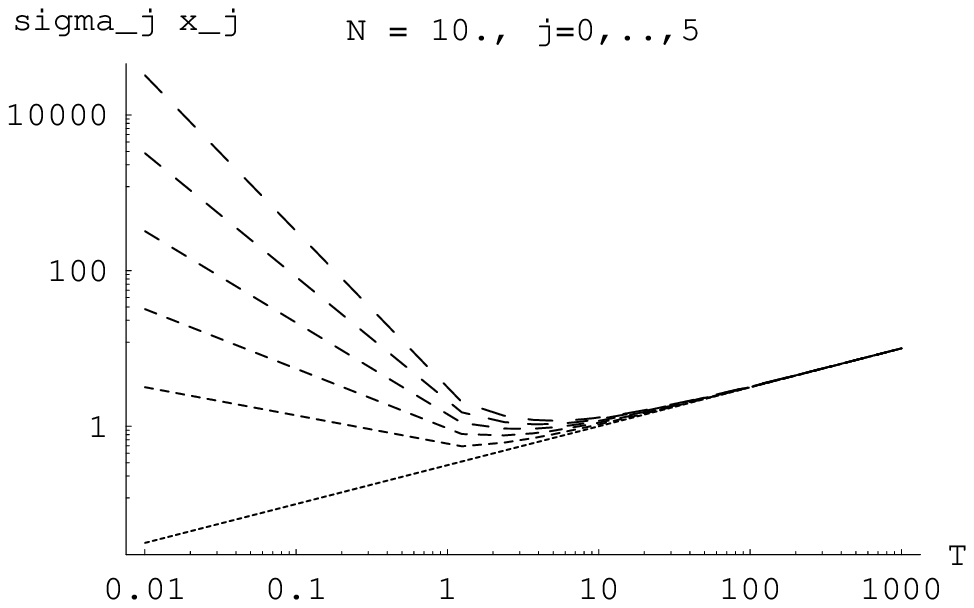,width=0.45\textwidth} 
\psfig{figure=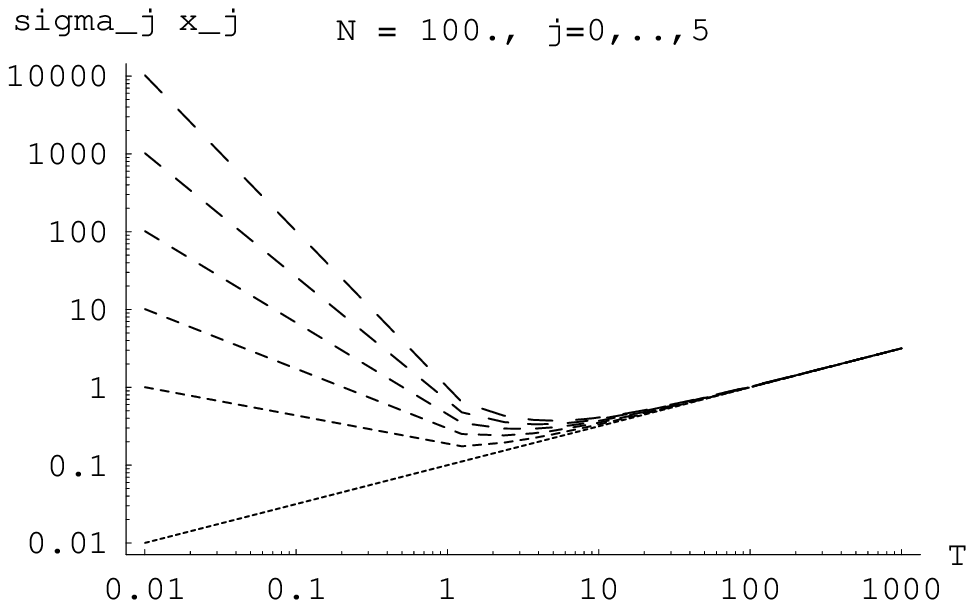,width=0.45\textwidth}}
\hbox{\psfig{figure=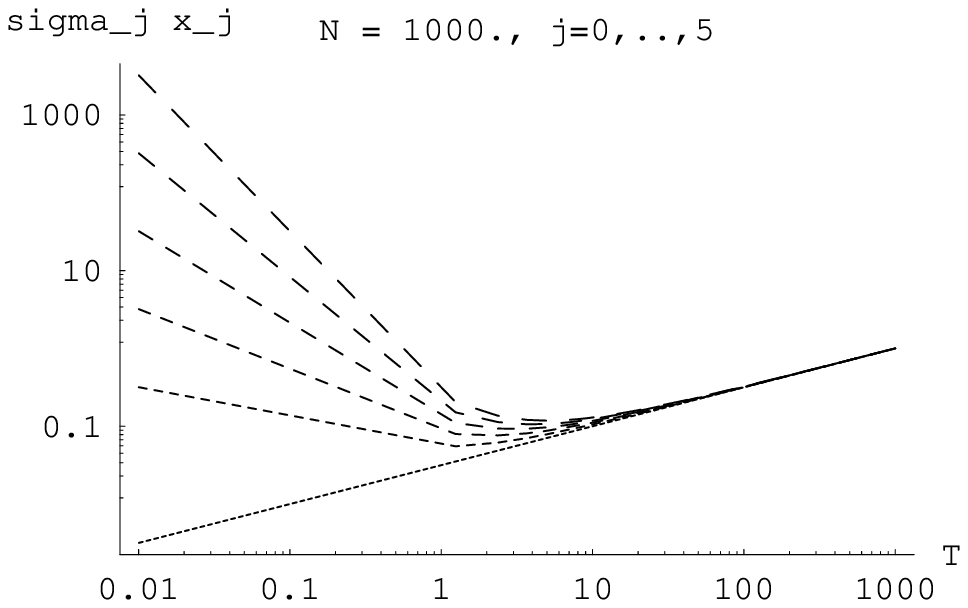,width=0.45\textwidth} 
\psfig{figure=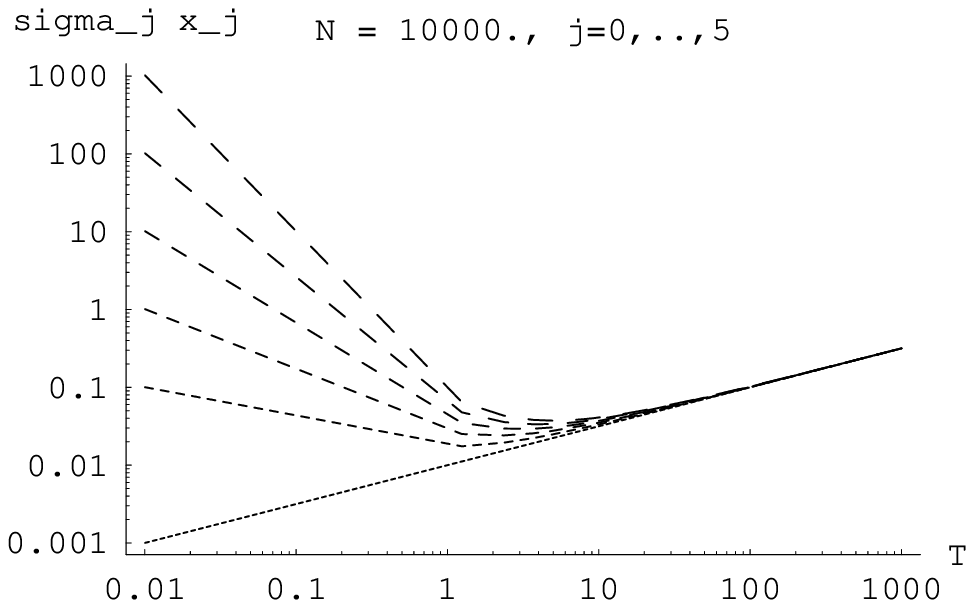,width=0.45\textwidth}}
\caption{The standard deviation of the occupation numbers of the energy levels in units of the mean occupation numbers 
as a function of temperature $T$ for $N=10,\dots,10^5$ and $j=0,\dots,5$ (with growing dash length).
At low and high temperatures the fluctuations are of the order of a couple of units of the mean occupation,
thus ``the system diverges''.\label{fig:VarianceFluctuations}}
\end{figure}

\begin{figure*}[tbh]
\hbox{
\psfig{figure=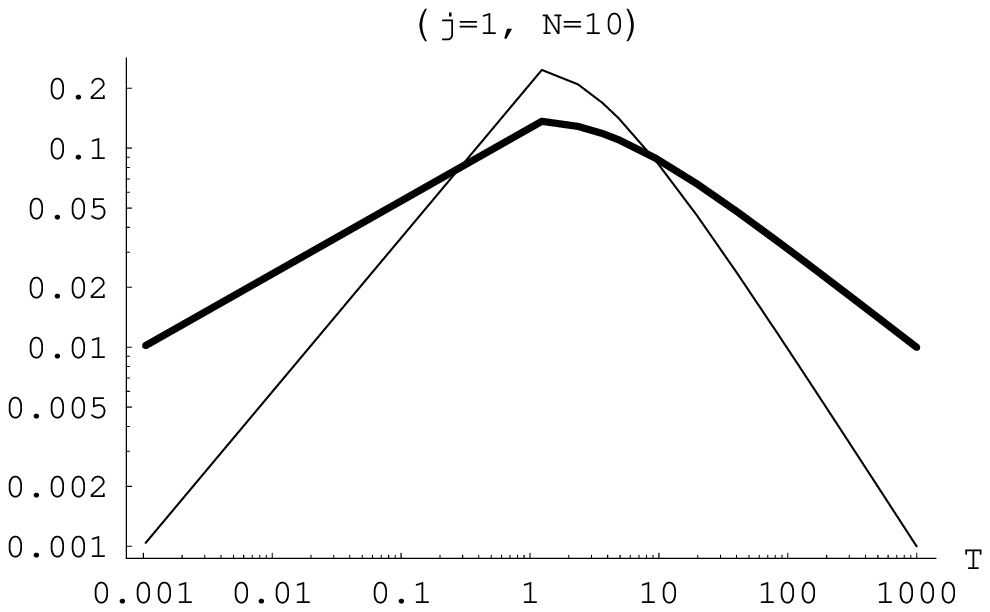,width=0.45\textwidth}
\psfig{figure=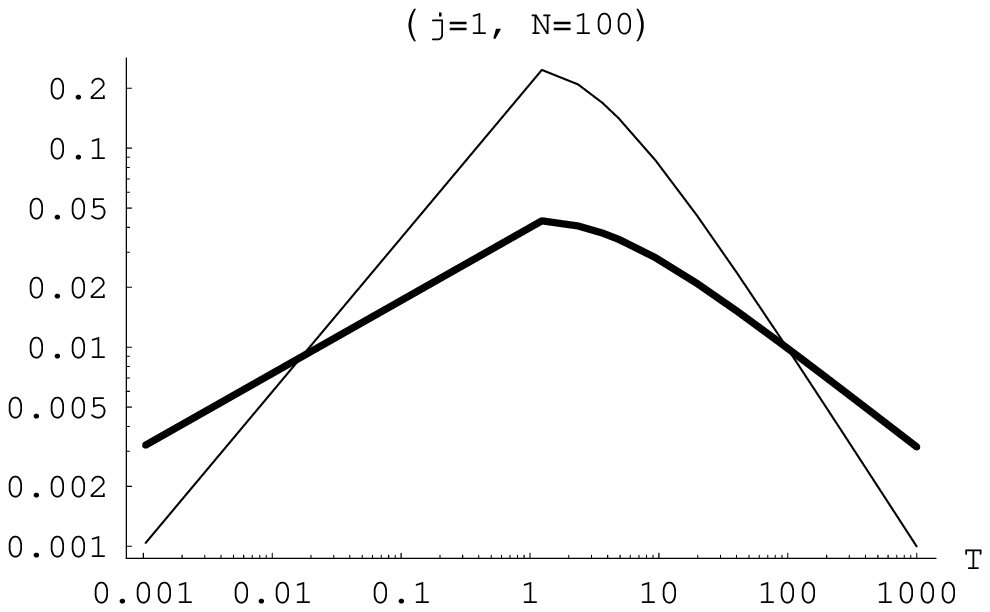,width=0.45\textwidth}}
\caption{The occupation density (thin line) of the first energy level, along with
its standard deviation (thick line) as a function of temperature. Here the number of particles reads
$N=10$ (left) and $N=100$ (right). We see that at low and at high temperatures
the standard deviation exceeds the mean occupation density and thus the system fluctuates strongly.
\label{fig:OccupationvsStds}}
\end{figure*}

To quantify the fluctuations we look at the variances of the occupation densities, which from (\ref{eq:Moments6}) read:
\begin{eqnarray}
\sigma_j^2 &=& \left<x_j^2\right> - \left<x_j\right>^2 
\nonumber \\ 
&=& 1_{j\le M}  \frac{C^{M-j+N-2}_{N-2}}{C^{M+N-1}_{N-1}} 
\left(\frac{1}{N} -  \frac{C^{M-j+N-2}_{N-2}}{C^{M+N-1}_{N-1}} \right)
\nonumber \\ 
&& {} + 1_{2j \le M} \frac{(N-1)}{N} \frac{C^{M-2j+N-3}_{N-3}}{C^{M+N-1}_{N-1}}
\label{eq:VarianceFluctuations} \\
&\mathop{=}_{N\rightarrow \infty} &
\frac{1}{N} \left<x_j\right>(1 - \left<x_j\right>)  =
\frac{1}{N} \frac{T^j}{(T+1)^{j+1}}
\left(1 - \frac{T^j}{(T+1)^{j+1}} \right)
\label{eq:VarianceFluctuations1} 
\end{eqnarray}
Equation (\ref{eq:VarianceFluctuations}) follows from the definition of the variance 
and equation (\ref{eq:VarianceFluctuations1}) follows from (\ref{eq:MthMomentLimit}) 
and (\ref{eq:MthMomentLimitResult}). For given $N$ and $j$, the variance is bounded from above as a function of $T$
and its maximal value, which occurs at the temperature  $T=j$, is given by:
\begin{equation}
(\sigma_{j}^{\mbox{max}})^2 = \frac{1}{N} 
\left<x_{j}^{\mbox{max}}\right> (1 - \left<x_{j}^{\mbox{max}}\right>)
\end{equation}
for $\left<x_{j}^{\mbox{max}}\right> = j^j/(j+1)^{j+1}$. Note that, for $j>0$, both 
the mean occupation density and the variance tend to zero at high and low 
temperatures, yet at different rates. Therefore we consider the ratio $\frac{\sigma_j}{\left<x_j\right>}$. We get:
\begin{equation}
\frac{\sigma_j}{\left<x_j\right>} = \frac{1}{\sqrt{N}} \sqrt{\frac{1- \left<x_j\right>}{\left<x_j\right>}}
=\left\{
\begin{array}{ll}
\frac{1}{\sqrt{N}} T^{-\frac{j}{2}} & \mbox{if}\quad T\to 0 \\
\frac{1}{\sqrt{N}} T^{\frac{1}{2}} & \mbox{if}\quad T\to \infty
\end{array}
\right.
\label{eq:STDvsMean}
\end{equation}
Thus, in the limit of high and low temperatures\footnote{As $T \to 0$ both $\left<x_j\right>$ and $\sigma_j$ approach zero for $j > 0$, but $\left<x_j\right>$ does so at a faster rate.}, the fluctuations prevail, the system diverges and the equilibrium distribution does not exist, contrary to the traditional view of classical statistical mechanics \cite{Huang2}.
We plot the standard deviation of the occupation densities in units of their mean occupation densities as a function of the temperature for different values of $N$ and 
$j$ in Figure \ref{fig:VarianceFluctuations}.
In addition, we plot the occupation density and its standard deviation for $j=1$ separately in Figure
\ref{fig:OccupationvsStds}.

We conclude this section by computing the distributions of the occupation densities, meaning we compute the likelihood that the number of particles $N_j$ at the $j^{th}$ level equals $n_j$. We get:
\begin{eqnarray}
\lefteqn{C^{M+N-1}_{N-1} P\left(N_j =\tilde{n}_j\right)}\nonumber \\
&&= 
\sum\limits_{\sum\limits_{q=0}^M n_q=N} \frac{N!}{\prod\limits_{q=0}^M n_q!} \delta_{M,\sum\limits_{q=0}^M q n_q} \delta_{n_j,\tilde{n}_j}
\label{eq:PDF0} \\
&&=
\left.\left.
\frac{1}{M!} \frac{d^M}{d z_1^M} \frac{1}{\tilde{n}_j!} \frac{d^{\tilde{n}_j}}{dz_2^{\tilde{n}_j}}
\left(\frac{1-z_1^{M+1}}{1-z_1} + z_1^j (z_2-1)\right)^N
\right|_{z_1=0}
\right|_{z_2=0}
\label{eq:PDF1} \\
&&=
\left.
\frac{1}{\tilde{n}_j!}
\frac{d^{\tilde{n}_j}}{dz_2^{\tilde{n}_j}}
\sum\limits_{q=0}^{N-1} 1_{q j\le M} (z_2-1)^q \alpha_q^{(M,j,N)} + \delta_{N j,M} (z_2-1)^N
\right|_{z_2=0}
\label{eq:PDF2} \\
&&=
\sum\limits_{q=\tilde{n}_j}^{N-1} 1_{q j\le M} C^q_{\tilde{n}_j} (-1)^{q-\tilde{n}_j} 
C^N_q C^{M-q j + N-1-q}_{N-1-q} + \delta_{N j,M} C^N_{\tilde{n}_j} (-1)^{N-\tilde{n}_j}
\label{eq:PDF3}
\end{eqnarray}
In the first equality in (\ref{eq:PDF1}) we repeated the manipulations from (\ref{eq:Microstates1})-(\ref{eq:Microstates2}), meaning we 
have inserted the integral representation of delta functions, performed the sum using the multinomial expansion formula and finally evaluated
the complex integrals over a pair of unit circles using the residue formula.
In (\ref{eq:PDF2}) we have computed the derivative by $z_1$ using equation (\ref{eq:PowerofSum}) and
in (\ref{eq:PDF3}) we computed the derivative by $z_2$ using derivatives of elementary functions.

\begin{figure*}[tbh]
\hbox{
\psfig{figure=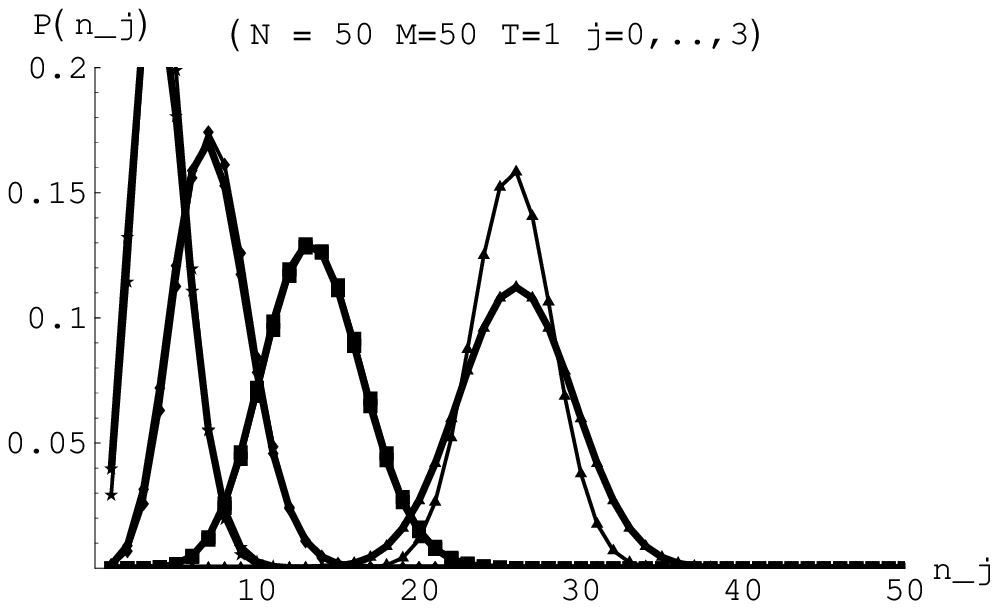,width=0.45\textwidth}
\psfig{figure=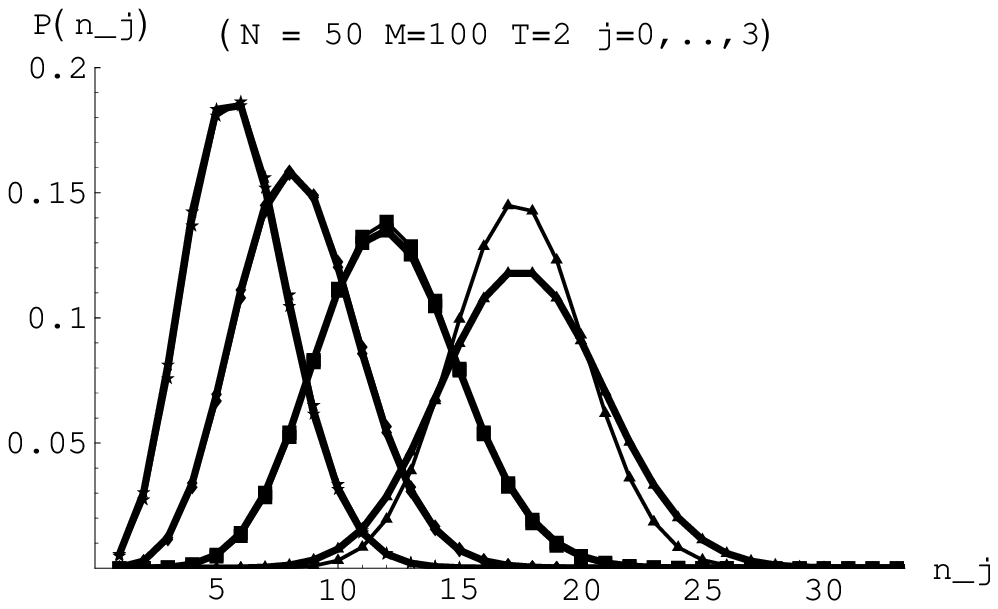,width=0.45\textwidth}}
\hbox{
\psfig{figure=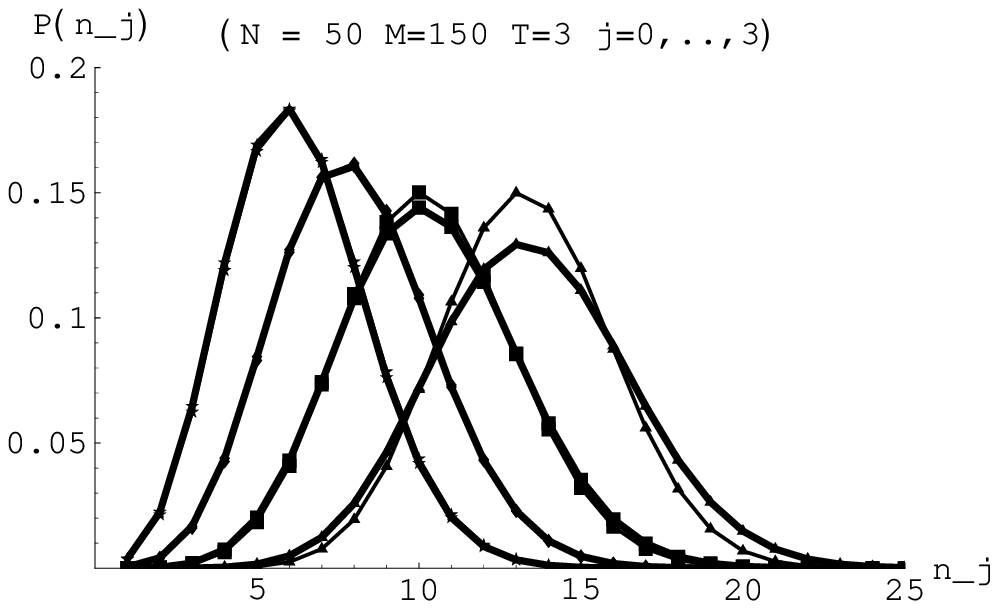,width=0.45\textwidth}
\psfig{figure=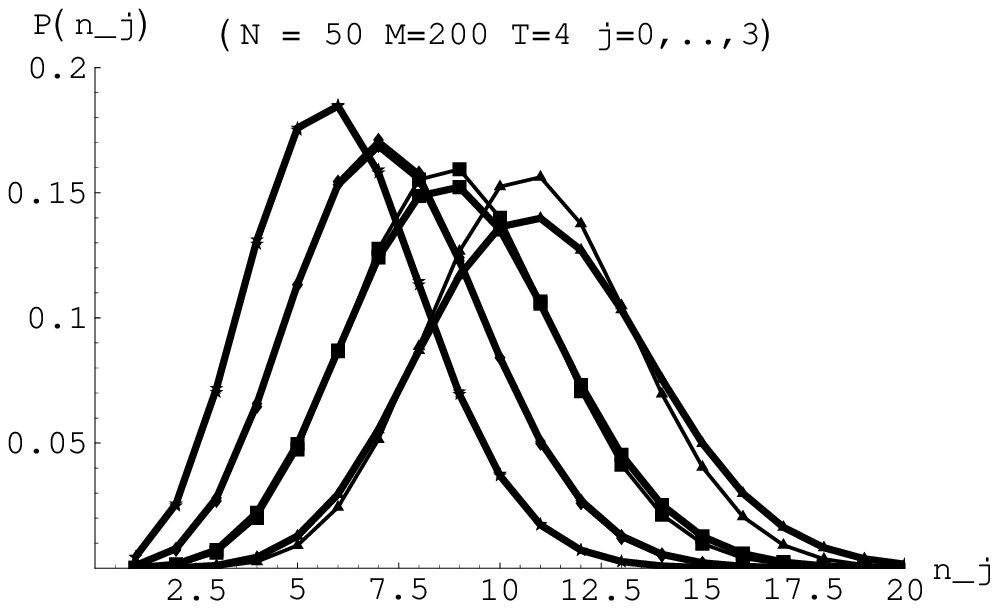,width=0.45\textwidth}}
\caption{The probability distributions, $P(n_j), $ of the occupations numbers of energy levels $j$ for both the exact formula (thin line) and the formula derived in the thermodynamic limit (thick line).
 Here $j=0,\dots,3$ (from right to the left) 
and $N=50$ and the value of $T$ in the plots clockwise from top left is $T:=M/N = 1,2,3,4$ respectively.\label{fig:PDFOccupation}}
\end{figure*}

\begin{figure*}[tbh]
\hbox{\psfig{figure=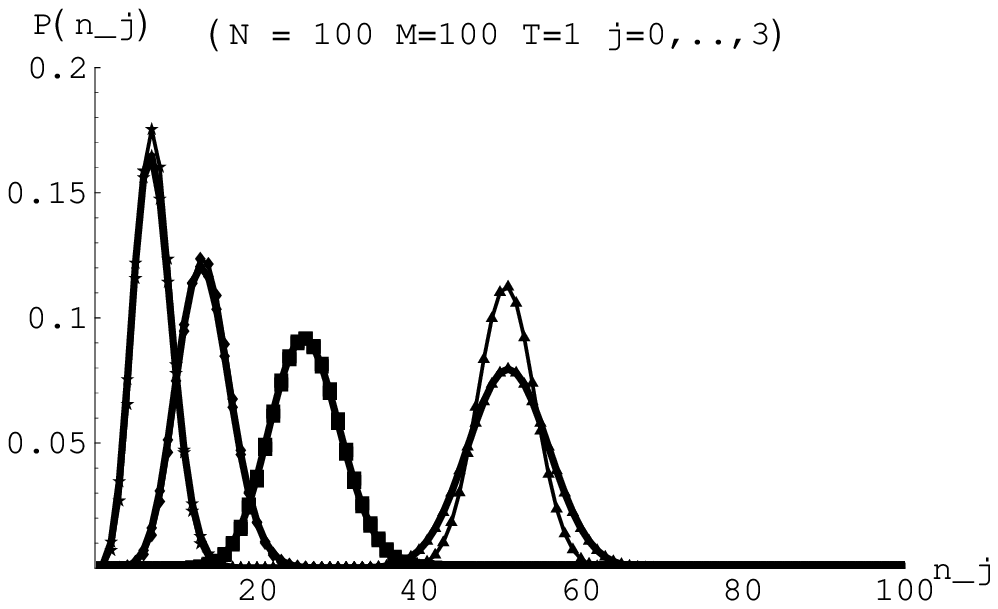,width=0.45\textwidth}
\psfig{figure=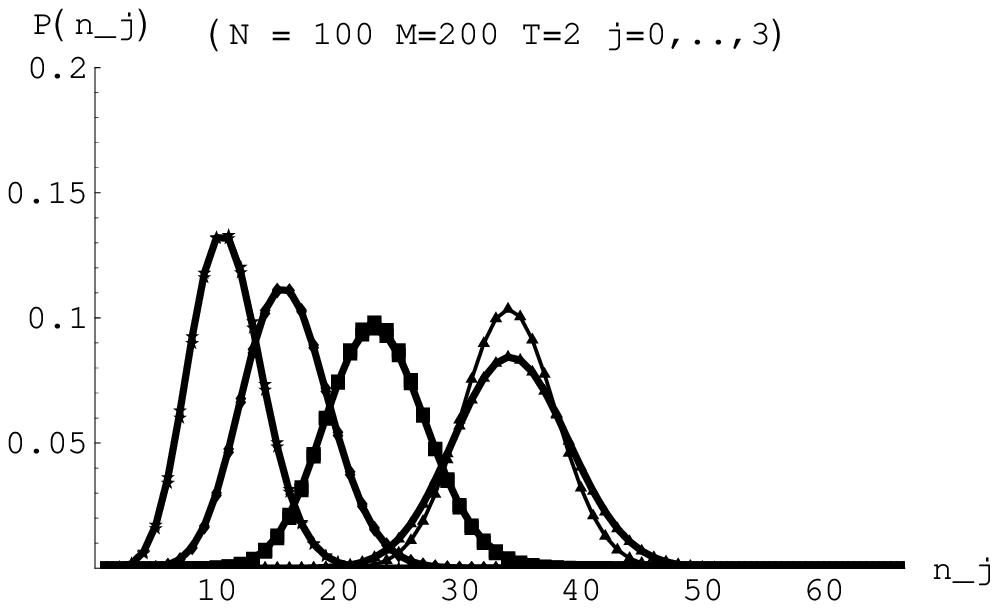,width=0.45\textwidth}}
\hbox{\psfig{figure=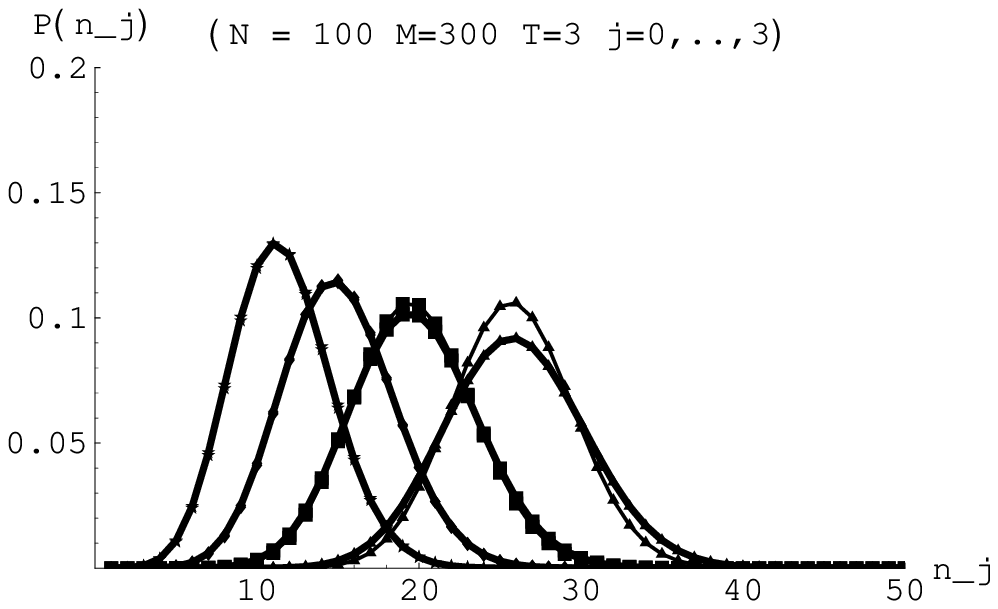,width=0.45\textwidth}
\psfig{figure=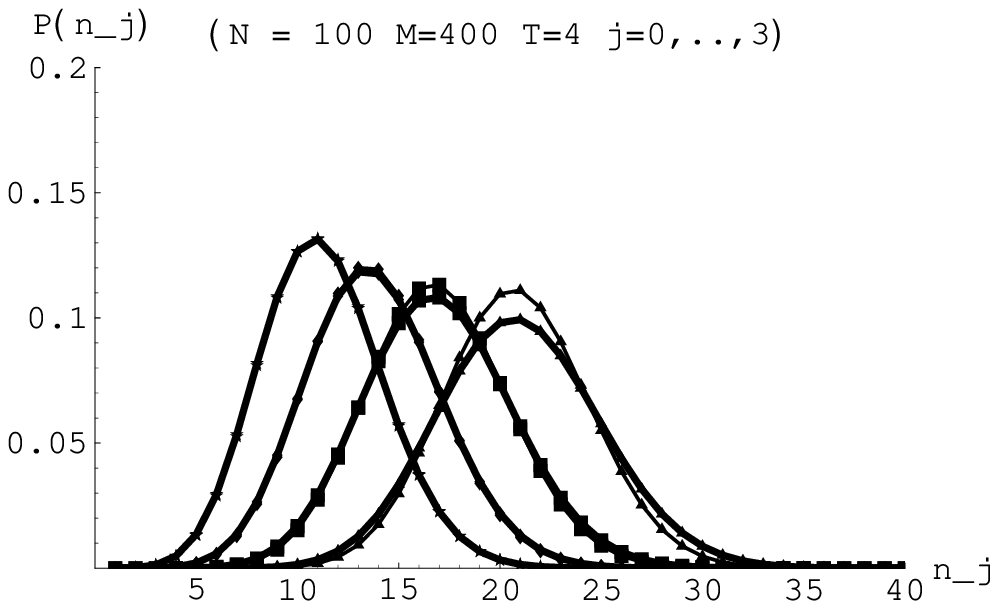,width=0.45\textwidth}}
\caption{Same as in Figure \ref{fig:PDFOccupation} but now have $N=100$.\label{fig:PDFOccupation1}}
\end{figure*}

For $j < T$ the large $N$ limit of the distributions of the occupation densities reads:
\begin{eqnarray}
\lim_{N\to \infty} P\left(N_j =\tilde{n}_j\right) &=& \sum\limits_{q=n_j}^N C^q_{\tilde{n}_j} (-1)^{q-\tilde{n}_j} C^N_q \left<x_j\right>^q 
\nonumber\\
&=& C^N_{\tilde{n}_j} \left<x_j\right>^{\tilde{n}_j} (1 - \left<x_j\right>)^{N-\tilde{n}_j} 
\label{eq:LargeNLimitPDF} \\
\nonumber  
&\simeq &  \mbox{Normal}(N \left<x_j\right>, N \left<x_j\right> (1-\left<x_j\right>))(\tilde{n}) \\
&:=& \frac{1}{\sqrt{2\pi N \left<x_j\right> (1-\left<x_j\right>)}} 
\exp\left(-\frac{(\tilde{n}- N \left<x_j\right>)^2}{2 N \left<x_j\right> (1-\left<x_j\right>)}\right)
\label{eq:LargeNLimitPDF1}
\end{eqnarray}
Equation (\ref{eq:LargeNLimitPDF}) follows from the fact that the large $N$ limit of the last binomial factor
on the right-hand side in (\ref{eq:PDF3}) reads $\left<x_j\right>^q$ and equation (\ref{eq:LargeNLimitPDF1})
follows from the normal approximation to the binomial distribution. 
We plot both the exact result and the large $N$ limit 
in Figures \ref{fig:PDFOccupation} and \ref{fig:PDFOccupation1} 
for several values of $N$ and $M$.

\begin{figure*}[tbh]
\hbox{
\psfig{figure=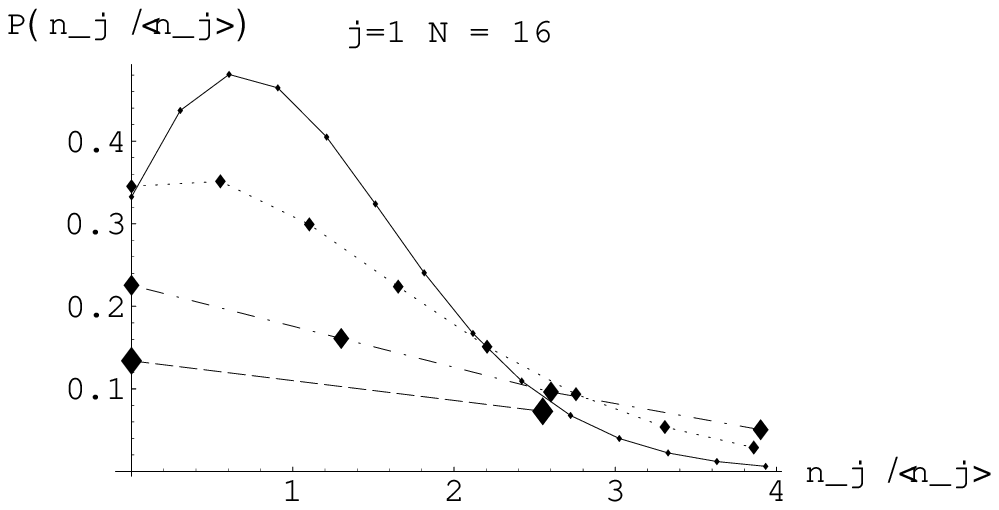,width=0.45\textwidth}
\psfig{figure=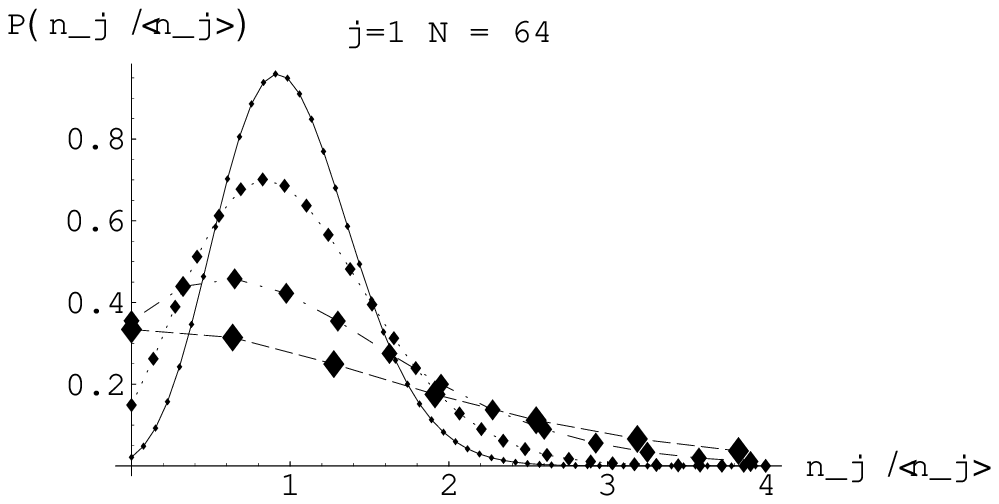,width=0.45\textwidth}}
\hbox{
\psfig{figure=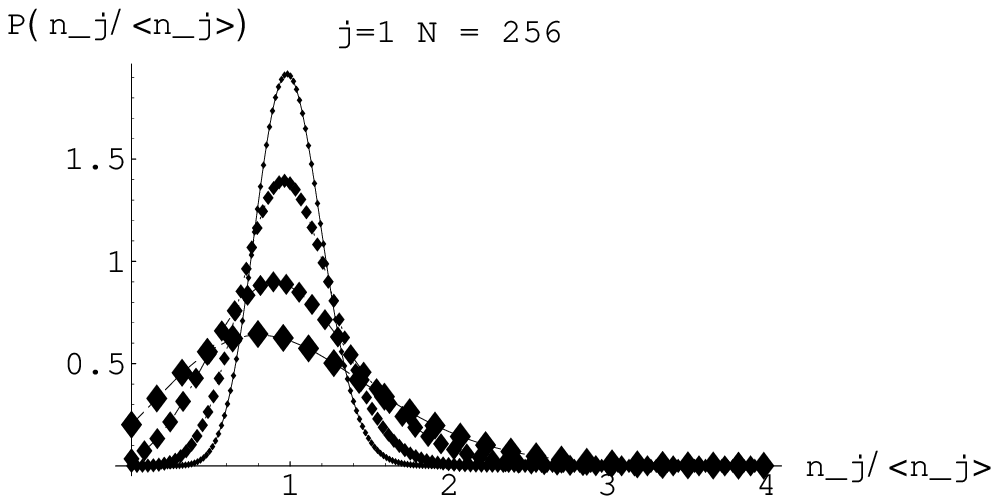,width=0.45\textwidth}
\psfig{figure=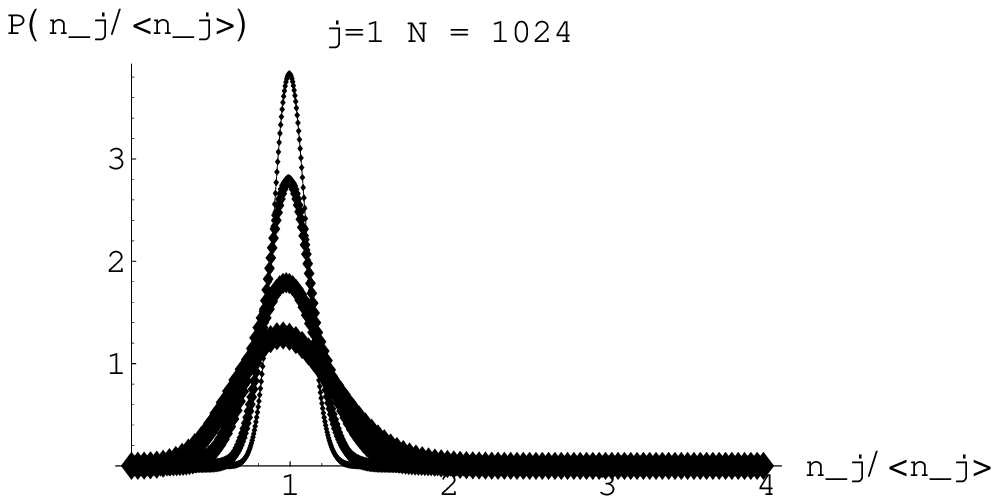,width=0.45\textwidth}}
\caption{The probability distribution of the occupation number of level $j=1$ for temperatures $T=10,20,50,100$ and for system sizes $N=16,64,256,1024$. For every finite $N$ the system ``diverges'' when the temperature goes to infinity.\label{fig:Fluctuations}}
\end{figure*}

Thus the distributions of the occupation densities, 
$x_j = n_j/N$, conform to Gaussians with means $N \left<x_j\right>$ and standard deviations 
$1/\sqrt{N} \sqrt{\left<x_j\right> (1-\left<x_j\right>)}$ 
in accordance with the results from \cite{Huang1}, for example.
However, for every finite $N$ if the temperature $T$ is high enough the occupation number in units of the mean occupation number has a distribution whose ``flatness'' or ``half width' is arbitrarily high.
In other words, for every finite $N$ the amount of fluctuations diverges when the temperature gets very large.
We depict that statement in Figure \ref{fig:Fluctuations}.

\subsubsection{A Note on Units}

In this paper, we call the ratio $\frac{M}{N}$ the ``temperature'' and discuss how high values of this quantity can lead to more fluctuations in the distribution. Equation~(\ref{eq:temp_def}) relates this ratio to the absolute temperature and the ``energy level spacing'', $\epsilon$, for the ideal gas system. 
But when is this ratio ``high''? That is, at what absolute temperature does this occur? 
Suppose we take the case where $M = 1000 N$, and we consider $\epsilon$ as some sort of precision limit on measurement e.g. $\sim 0.1\mbox{eV } \simeq 1.6 \times 10^{-20}$J. Note that this is the amount of energy in an infrared light quantum and, as such, is measurable by spectroscopical methods. From~(\ref{eq:temp_def}) this gives an absolute temperature of about 770,000 K which is certainly hotter than anything naturally occurring on Earth, though would be easily exceeded in most stars. Thus our analysis would not be useful for physical systems at room temperature although may be of interest for astrophysics.

However, suppose now we consider a financial system, and let the ``energy quantum'' represent now a unit of money, e.g. one dollar (with the Boltzmann constant being suitably redefined or discarded). A typical financial system, where the number of market participants is much less than the amount of money involved, could give values for the ratio of $10^{4}$, or much higher. The considerations in our paper would therefore take on greater importance for these systems.

\subsubsection{Multi-point (Macrostate) Probability Distribution Function} 

We now generalize the calculations in (\ref{eq:PDF0})-(\ref{eq:PDF3}) to obtain the multivariate 
probability function. We take $p=1,\dots,M+1$ and an ascending integer sequence
$0\le j_1 < \dots < j_p \le M$ and
we define the $p$-variate probability function
$P\left(\bigcap\limits_{s=1}^p \left( N_{j_s}=\tilde{n}_{j_s}\right) \right)$ as the probability that
we find $\tilde{n}_{j_s}$ particles at energy level $j_s$ for $s=1,\dots,p$.
The function is given in (\ref{eq:PDFMultiVarMainText3}). We have:
\begin{eqnarray}
\lefteqn{C^{M+N-1}_{N-1} P\left(\bigcap\limits_{s=1}^p \left( N_{j_s}=\tilde{n}_{j_s} \right) \right)=}
\nonumber \\&&
\!\!\!\!\!\!\!\!\!\!\!\!\!
\sum\limits_{q=0}^{N-1}
1_{J^{\vec{s}}_q\le M}
\left(
\prod\limits_{l=1}^p C^{m_l^{(q)}}_{\tilde{n}_{j_l}} (-1)^{m_l^{(q)} - \tilde{n}_{j_l}} 1_{\tilde{n}_{j_l} \le m_l^{(q)}}
\right)
C^N_q C^{M - J^{\vec{s}}_q+N-1-q}_{N-1-q}
\nonumber \\
&& \mbox{} + \delta_{J^{\vec{s}}_q,M}
\left(\prod\limits_{l=1}^p C^{m_l^{(N)}}_{\tilde{n}_{j_l}} (-1)^{m_l^{(N)} - 
\tilde{n}_{j_l}} 1_{\tilde{n}_{j_l} \le m_l^{(N)}}\right)
\label{eq:PDFMultiVarMainText3}
\end{eqnarray}
subject to:
\begin{equation}
\nonumber
m_r^{(q)} = \sum\limits_{l=1}^q \delta_{r,s_l} \mbox{for }r=1,\dots, p \quad\quad\mbox{and}\quad\quad 
 m_r^{(N)} = \sum\limits_{l=1}^N \delta_{r,s_l} \mbox{for }r=1,\dots,p  .
\end{equation}

Here, we have defined:
\begin{equation} 
J^{\vec{s}}_q := \sum\limits_{l=1}^q j_{s_l}
\label{eq:CapJs}
\end{equation}
Note that for every $q=0,\dots,N-1$ the expression on the right-hand side is summed over integer grid-points
$\left(s_l\right)_{l=1}^q$
of a $q$-dimensional hypercube of side length $p$ i.e.\ $\left(s_l\right)_{l=1}^q \in \lbrace 1, 2, \ldots p \rbrace$. As such the expressions in parentheses contain $p^q$ terms.
The expression on the right-hand side depends on $\left(m_l^{(q)}\right)_{l=1}^p$. These numbers
count how many coordinates of the  grid-point are equal to $l=1,\dots,q$. 
Therefore the $p$-variate probability function is invariant under permutations of its arguments.
In addition, the identity $\sum\limits_{l=1}^p m_l^{(q)} = q$ holds true.
The proof is in~\ref{AppC}.

In the case $p=1$ we have $s_1=\dots=s_N=1$ and $m_1^{(N)} = N$ and 
$j_{s_1} = \dots = j_{s_q} = j_1 = j$, the expressions in parentheses reduce to one term only, 
and we clearly retrieve the probability function in (\ref{eq:PDF3}).

Expression (\ref{eq:PDFMultiVarMainText3}) is cumbersome and does not provide much insight in to the problem.
Thus, in order to better understand the expression, we compute its large-$N$ limit. 
We prove that the $p$-variate probability function is normalized to unity in~\ref{AppE}.
In addition we take $j_s = s-1$ for $s=1,\dots,p$., i.e.\ we are interested in the $p$ lowest energy levels.
The result is a multinomial distribution with likelihoods of individual trials ${\mathfrak p}_l$ given 
in (\ref{eq:IndivProbs}). We have:
\begin{equation}
P\left(\bigcap\limits_{s=1}^p \left(N_{j_s}=\tilde{n}_{j_s}\right)\right)=
\frac{N!}{(N- \left|\vec{\tilde{n}}\right|)!(\prod\limits_{l=1}^p \tilde{n}_{j_l}!)} \prod\limits_{l=1}^{p+1} {\mathfrak p}_l^{\tilde{n}_{j_l}}
\label{eq:LargeNLimMultVariateFinalI} 
\end{equation}
where 
\begin{equation}
{\mathfrak p}_l := 
\left\{
\begin{array}{rr}
(\frac{T}{T+1})^l\frac{1}{T} & \mbox{if $l=1,\dots,p$}\\
(\frac{T}{T+1})^p & \mbox{if $l=p+1$}
\end{array}
\right.
\label{eq:IndivProbs}
\end{equation}
The proof is in~\ref{AppD}.

\paragraph{Specific cases}
\hspace{2 pt} \newline
In the case $p=1$ the expression (\ref{eq:LargeNLimMultVariateFinal3}) reduces to the binomial distribution with mean $N \left<x_0\right>$
and variance $N \left<x_0\right>(1- \left<x_0\right>)$ in accordance with equation (\ref{eq:LargeNLimitPDF}).
In the case $p=M+1$, from (\ref{eq:LargeNLimMultVariateFinal3}), (\ref{eq:IndivProbs}) and the conservation laws 
(\ref{eq:ConservationLaws}) and by using the Stirling approximation
to the binomial coefficient, we get:
\begin{eqnarray}
\lefteqn{P\left(\bigcap\limits_{l=0}^M \left( N_{l}=\tilde{n}_{l}\right) \right)=
\frac{N!}{\prod\limits_{l=0}^M \tilde{n}_{l}!} 
\frac{T^M}{(T+1)^{N+M}}}
\nonumber \\&&=
\left(\frac{1}{\sqrt{2\pi}} \frac{1}{\sqrt{T (T+1)}} \right) 
\frac{N!}{\prod\limits_{l=0}^M \tilde{n}_{l}!} 
\frac{1}{C^{M+N-1}_{N-1}}
\label{eq:TotalFct}
\end{eqnarray}
Here we have used $n_{M+1} = N - \sum\limits_{i=0}^{M}n_i = 0$. 
The last line in (\ref{eq:TotalFct}) is, except for a multiplicative prefactor, equal to the number of 
microstates corresponding to one macrostate, divided by the total number of 
macrostates (compare (\ref{eq:Multiplicity}) and (\ref{eq:Microstates2})), which is what we would expect. 
The fact that the prefactor is not unity is due to the approximations made (large $N$ approximations, Stirling approximation etc.). 
This could be improved upon using higher orders but is not needed here.

\paragraph{The mean, the covariance and the correlation matrices}
\hspace{2 pt} \newline
Now take $p=M+1$.
In order to quantify the entire amount of fluctuations  we give the mean and the covariance matrix of the $(M+1)$-point distribution function.
The mean reads:
\begin{equation}
\left<n_l\right> = N {\mathfrak p}_{l+1} = N\left<x_l\right>
\quad\mbox{for $l=0,\dots,M$}
\label{eq:MeanMultivariate}
\end{equation}
in accordance with the mean of the univariate distribution, given in (\ref{eq:MthMomentLimitResult}). 
The covariance matrix reads:
\begin{eqnarray}
c_{l_1,l_2} &:=& \left<(n_{l_1}-\left<n_{l_1}\right>)(n_{l_2}-\left<n_{l_2}\right>)\right> \nonumber \\
&=&
\left\{
\begin{array}{rll}
-N {\mathfrak p}_{l_1+1} {\mathfrak p}_{l_2+1} &= -N \left<x_{l_1}\right>\left<x_{l_2}\right>& \mbox{if $l_1\ne l_2$} \\
N {\mathfrak p}_{l_1+1} (1 - {\mathfrak p}_{l_1+1}) &= N \left<x_{l_1}\right>(1-\left<x_{l_1}\right>)& \mbox{otherwise}
\end{array}
\right.
\quad\mbox{}
\label{eq:CovarMultivariate}
\end{eqnarray}
for $l_1,l_2=0,\dots,M$.
in accordance with (\ref{eq:VarianceFluctuations1}).
The results (\ref{eq:MeanMultivariate}) and (\ref{eq:CovarMultivariate}) follow from the fact that 
the multivariate distribution is a multinomial distribution.

\begin{figure*}[tbh]
\hbox{
\psfig{figure=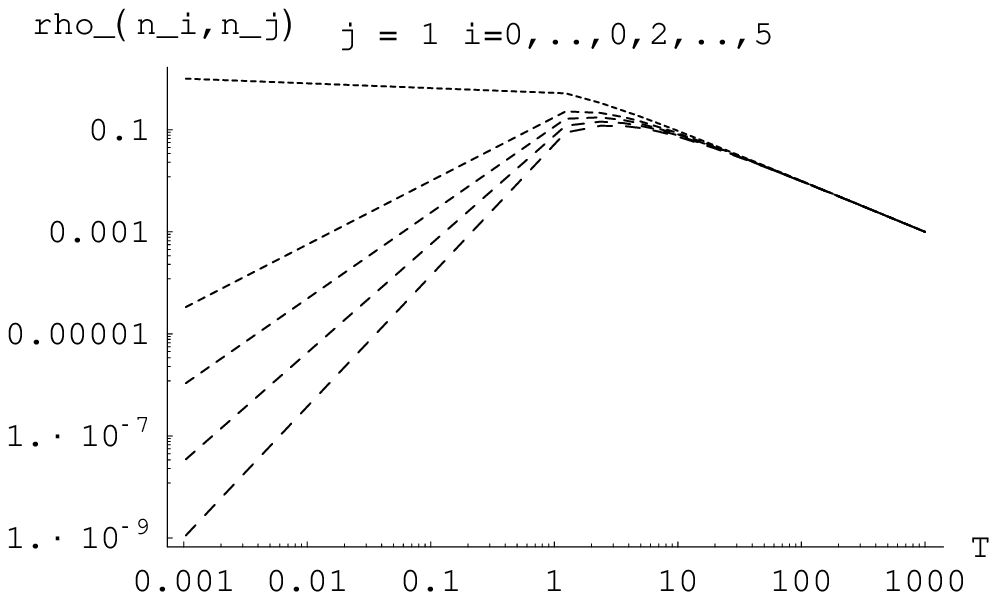,width=0.45\textwidth}
\psfig{figure=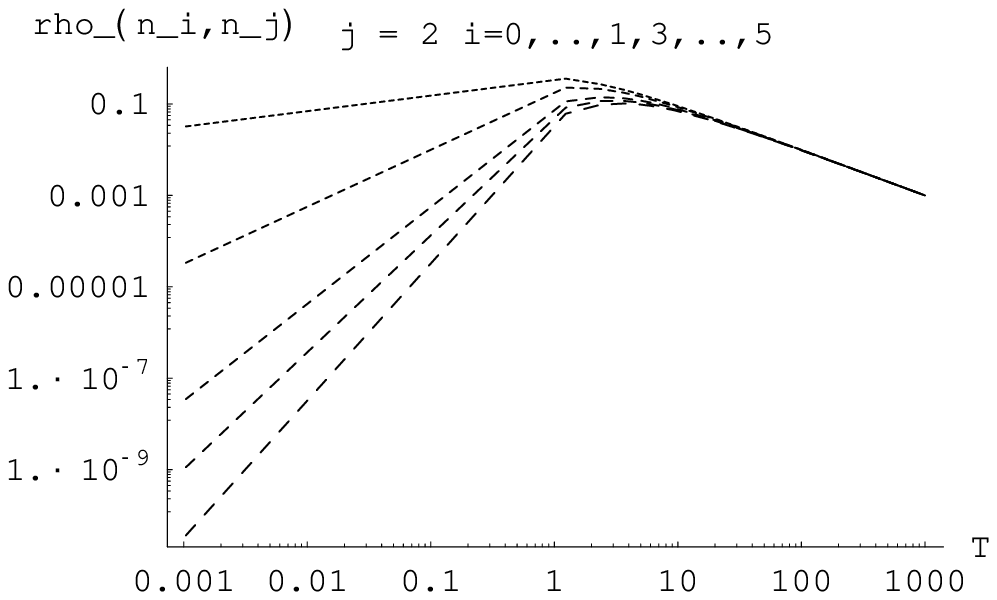,width=0.45\textwidth}
}
\hbox{
\psfig{figure=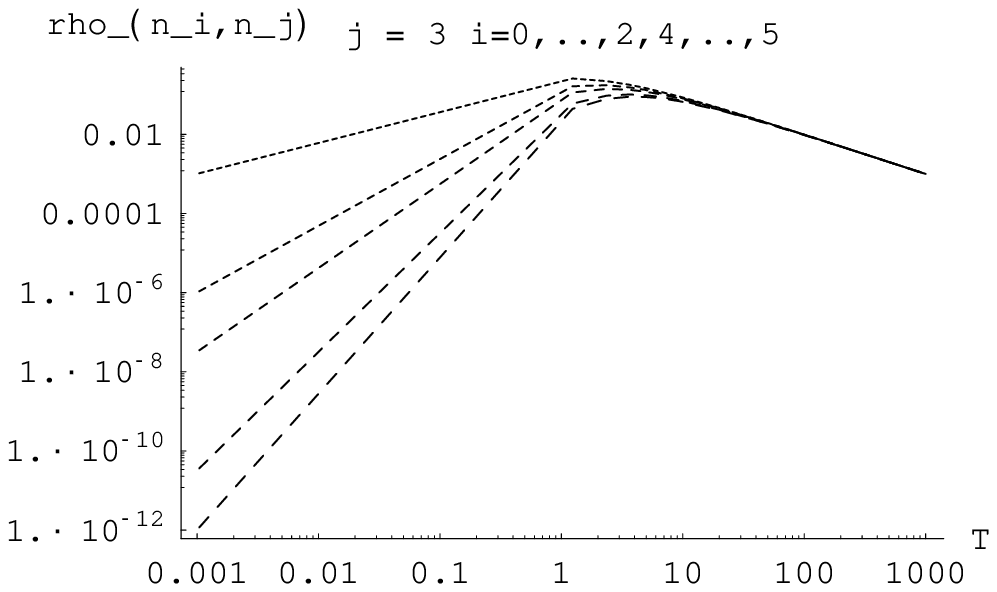,width=0.45\textwidth}
\psfig{figure=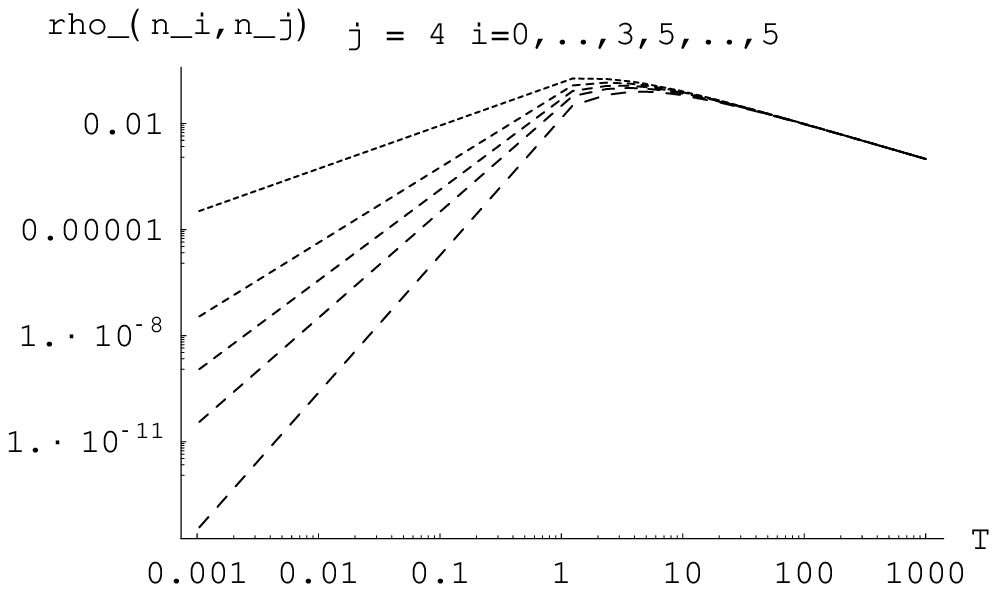,width=0.45\textwidth}
}
\caption{The modulus of correlations between occupation numbers at different levels $i=0,\dots,5$ (with increasing dash length) and $j$ for $j=1,\dots,4$.
Occupations of two different levels can be roughly treated as independent from each other unless $i,j=0,1$ or $i,j=1,0$
and temperatures are close to unity.
\label{fig:PearsonCorrel}}
\end{figure*}

From (\ref{eq:MeanMultivariate}) and (\ref{eq:CovarMultivariate}) we compute the Pearson correlation coefficient $\rho_{n_i,n_j}$
between occupation numbers
at levels $i \ne j$.
\begin{eqnarray}
\rho_{n_i,n_j} &:=& \frac{\left<(n_{i}-\left<n_{i}\right>)(n_{j}-\left<n_{j}\right>)\right>}{\sigma_j \sigma_j}
\nonumber\\
&=&
-\sqrt{\frac{\left<x_i\right>\left<x_j\right>}{(1-\left<x_i\right>)(1-\left<x_j\right>)}}
\nonumber \\
&=&
\left\{
\begin{array}{ll}
-T^{(i+j)/2} & \mbox{for $T\rightarrow 0$ and $i,j>0$} \\
-T^{-1} & \mbox{for $T\rightarrow \infty$}
\end{array}
\right.
\end{eqnarray}
We can see that occupation numbers of  any two different levels  are anticorrelated.
 The occupation numbers become uncorrelated once temperatures become high or low
and a maximal correlation is attained at some intermediate temperature (see Figure \ref{fig:PearsonCorrel}).
This is what one would intuitively expect, since at both high and low temperatures the mean occupations tend to zero and as such can be treated as independent
from each other.

\begin{figure*}[tbh]
\hbox{
\psfig{figure=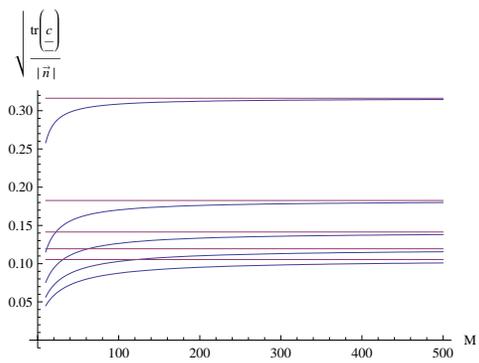,width=0.45\textwidth}
}
\caption{The total amount of fluctuations of the vector of the occupation numbers, graphed as a function of $T$, for $N=10, 30, 50, 70, 90$. The amount of fluctuations were expressed as the trace of the covariance matrix of the occupation numbers in units of the $L^1$ norm of the vector of mean occupation numbers. 
\label{fig:flucs_high_T_limit}}
\end{figure*}

\paragraph{``The total amount of fluctuations'' in units of the mean occupation.}
\hspace{2 pt} \newline
We conclude this section by investigating how strong is the effect of fluctuations on the whole vector of occupation numbers
rather than only on one particular occupation number, which was the problem that we explored in (\ref{eq:STDvsMean}).
Here we need to use some measures associated with the covariance matrix and the vector of mean occupation numbers.
We choose to compare the square root of the trace 
of the covariance matrix to the $L^1$ norm of the vector of mean occupation numbers.
Note that this is a natural generalisation of the variance for multivariate distributions, 
i.e.\ it is of the form: $\frac{\left< \| \vec{Y}- \left< \vec{Y}\right> \|^2 \right>}{|\left< \vec{Y} \right>|}$, where $\| . \|$ and $| . |$ are the $L^2$ and $L^1$ norms respectively and $\vec{Y}$ is the random vector.
We define $\underline{\underline{c}} := \left(c_{i,j}\right)_{i,j=0}^M$ and $\vec{n} := \left(\left<n_i\right>\right)_{i=0}^M$, and make the substitution ${\mathfrak x} = \frac{T}{T+1}$.
Using (\ref{eq:MeanMultivariate}) and (\ref{eq:CovarMultivariate}), we calculate the following:
\begin{eqnarray}
\frac{\sqrt{\mbox{tr}\left(\underline{\underline{c}}\right)}}{\left|\vec{n}\right|} &=&
\sqrt{\frac{{\mathfrak x}}{(1+{\mathfrak x})}}
\frac{\sqrt{2 - {\mathfrak x}^M - {\mathfrak x}^{M+1} + {\mathfrak x}^{2M+1} - {\mathfrak x}^{2M+2}}}
{\sqrt{N} \left( 1-{\mathfrak x}^{M+1} \right)}
\label{eq:multipoint_flucs}\\
&=&
\frac{1}{\sqrt{N}}
\left\{
\begin{array}{ll}
\frac{1}{\sqrt{1 - e^{-N}}}  & \mbox{if $T\rightarrow \infty$} \label{eq:total_flucs_high_T_limit} \\
\sqrt{T} & \mbox{if $T\rightarrow 0$}\label{eq:total_flucs_low_T_limit}
\end{array}
\right.
\end{eqnarray}
We examine the approach to the high temperature limit more closely by graphing (in figure \ref{fig:flucs_high_T_limit}) the function in equation (\ref{eq:multipoint_flucs}) against $T$, for different values of $N$. 
We can see that, for fixed $N$, the total amount of fluctuations increase with $T$ towards the constant value given in equation (\ref{eq:total_flucs_high_T_limit}), as opposed to the univariate case in equation~(\ref{eq:STDvsMean}) where the ``amount of fluctuations'' diverged as $T\to \infty$. 
Therefore, the conclusion from (\ref{eq:STDvsMean}) still holds and the system does not converge on the Maxwell-Boltzmann distribution but instead  fluctuates around it except that the total amount of fluctuations asymptotically approaches a finite limit.

\subsubsection{Corrections to Fluctuations for Finite $N$}

We can see, from (\ref{eq:LargeNLimitPDF1}), that in the thermodynamic limit, $N\rightarrow \infty$, the amount of fluctuations present in the system
depends on the temperature $T$ and the total number of particles $N$. For finite systems, however, there 
are corrections to the amount of fluctuations which we specify now.
We note that the last term in (\ref{eq:PDF3}) is absorbed as the $q=N$ term into the sum and the 
sum over $q$ is performed by expressing the ratio of binomial factors as a Laplace transform by $q$ viz:
\begin{equation}
1_{q j\le M} \frac{C^{M + N-1-q(j+1)}_{N-1-q}}{C^{M+N-1}_{N-1}} =
\left< \theta^j \right> =: \int\limits_0^1 d\theta \omega_j(\theta) \theta^q
\label{eq:LaplaceTrafo}
\end{equation}
where the weight $\omega_j(\theta)$ reads:
\begin{equation}
\omega_j(\theta) = \int\limits_{\imath {\mathbb R}} \frac{dq}{2\pi \imath} \theta^{-(q+1)}
1_{q j\le M} \frac{C^{M + N-1-q(j+1)}_{N-1-q}}{C^{M+N-1}_{N-1}}
\label{eq:ComplexIntegral}
\end{equation}

Exchanging the integration in (\ref{eq:LaplaceTrafo}) with the sum over $q$ in (\ref{eq:PDF3}) we easily arrive at the
following result:
\begin{equation}
P\left(N_j =\tilde{n}_j\right) = 
C^N_{n_j} \int_0^1 d\theta\omega_j(\theta) \theta^{n_j} (1 - \theta)^{N-n_j}
\end{equation}

The integral in (\ref{eq:ComplexIntegral}) can be done analytically by using complex calculus, however, since 
we see in Figures \ref{fig:PDFOccupation} and \ref{fig:PDFOccupation1} that the corrections are not higher than a couple
of percent even for $N$ of the order of a couple of tens we do not see any motivation in deriving 
these results analytically. Thus, we conclude, the distribution of the number of particles on the $j^{th}$ level is a continuous linear 
superposition of Binomial distributions with mean $N\theta$ and variance $N\theta(1-\theta)$ and weights $\omega_j(\theta)$
for $\theta \in [0,1]$. In the thermodynamic limit the weight tends towards a Dirac delta function that picks out 
$\theta = \left<x_j\right>$ and we retrieve the result from equation~(\ref{eq:LargeNLimitPDF}).

\section{Conclusions} 
It has long been understood that
the state to which an isolated thermodynamic system is attracted to, 
as a result of the Law of Large Numbers (i.e.\ $N \to \infty$), becomes an exponential distribution in the thermodynamic limit.
In this document however, we have shown that for every finite $N$ (large enough for some of the formulae simplifications used in this document to be valid), there always exists some degree of fluctuations that depend on the temperature of the system.
The occupations of the energy levels fluctuate, and their distributions can be well approximated by normal curves with means $N \left<x_j\right>$ 
and variances $N \left<x_j\right> (1-\left<x_j\right>)$ (here $\left<x_j\right>$ is the mean occupation density of the $j^{\mbox{th}}$ energy level and is given in (\ref{eq:MthMomentLimit})). In fact for higher temperatures the fluctuations can be considerable relative to the mean occupation, as seen in Figure (\ref{fig:VarianceFluctuations}) and Figure (\ref{fig:OccupationvsStds}). Thus, under these conditions, one cannot speak about a static distribution of particles among the energy levels since in any finite fixed $N$ system, there is always some degree of fluctuation about the underlying mean distribution. 

In this work we used ``temperature'' in the sense of a specific energy (energy per particle). If we look at other systems, this temperature could be a proxy for any ``{\it mean quantity}'' (i.e.\ the amount per component, of a conserved quantity that is distributed among the components) in any finite $N$ thermodynamic-like system. This understanding that there are increasing degrees of disorder associated with an increasing mean quantity can be applied to any average quantity. 

Furthermore, several people~\cite{Dragulescu,Ferrero,Yuqing,Kleinert} have recently applied Boltzmann statistics to income distributions and the financial markets. We believe our work can help explain why financial markets are an economic equilibrium that is not static, but fluctuating. We believe this work has the potential to lead to a method for measuring the temperature (i.e.\ the amount of fluctuations) and  the rate of change of temperature (i.e.\ trending) in financial and property markets.

But possibly the most interesting thing of all is to think about what happens in a thermodynamic-like system, 
when the components of the system are not simple things like particles, but entities with some intelligence having 
the ability to adapt to their environment. Such adaptive entities would be capable of choosing to convert 
potential energy stores into kinetic energy (and vice versa) and as a result the overall system would effectively have the ability 
to ``{\it endogenously}''  alter its own temperature.  

It has long been understood that simple systems spontaneously move to a thermal equilibrium. 
In future work we hope to show that (as a result of competition and cooperation) ``{\it complex adaptive systems}'' 
gravitate to another type of equilibrium, an equilibrium somewhere between order and disorder, an equilibrium which maximizes 
survival by maximizing flexibility, an equilibrium which we might call a ``{\it structural}'' or 
``\underline{{\it organizational equilibrium}}''. 

\section*{Acknowledgements} We are grateful to Piotr Fronczak and Agata Fronczak and Stefan Thurner for bringing to our attention the
works of Evans and coworkers \cite{Evans} on the Fluctuation Theorem~\cite{Wiki}. 
Since this Theorem deals with deviations from the Second Law 
of Thermodynamics in classical dissipative systems the Theorem is closely related to our work that too deals with fluctuations 
at the thermodynamic equilibrium. We stress, however, that our work is based on different assumptions; 
we deal with a conservative system
at equilibrium rather than with a dissipative system moving towards equilibrium.

\appendix

\newcommand{\appsection}[1]{\let\oldthesection\thesection\renewcommand{\thesection}{Appendix \oldthesection}\section{#1}\let\thesection\oldthesection}

\appsection{}\label{AppA} 

Here we list some important identities that are used in this paper.
We believe that these identities are useful {\it per se} in analysis, partial differential equations,
and probability \& statistics.

\noindent{\bf The Sum of Powers Identity:}

\begin{equation}
\sum\limits_{l=0}^{t-1} l^n = 
\sum\limits_{k=0}^{n+1} t^{n-k+1} \frac{n!}{(n-k+1)!} c_k
\label{eq:SumIdentity}
\end{equation}
where 
\begin{equation}
c_k := \sum\limits_{p=0}^k (-1)^p 
\mathop{\sum\limits_{n_1+\dots+n_p = p+k}}_{n_1,\dots,n_p \ge 2} \frac{1}{\prod\limits_{q=1}^p n_q!}
=\left(1,-\frac{1}{2},\frac{1}{12},-\frac{1}{8},\dots\right) 
\label{eq:SumofPow} 
\end{equation}
for $k=0,\dots,n+1$. The identity (\ref{eq:SumofPow}) can be proven by using a trick 
\mbox{$\left.d^n_{\log(t)^n} t^l \right|_{t=1} = l^n$} for \mbox{$n,l \in {\mathbb N}$},
changing the order of differentiation and summation, re-summing the resulting geometric series,
and differentiating the result using the chain rule of differentiation.

\noindent{\bf The Combinatorial Identity:}

\begin{equation}
\sum\limits_{1\le l_1 < \dots < l_s \le 4-n}
\prod\limits_{j=0}^s C^{p_{j+1} - p_j-1}_{l_{j+1} - l_j-1}
=
C^{4-s}_{4-n-s} 
\label{eq:BinomialFactorIdentity}
\end{equation}
with $l_0=p_0=0$ and $l_{s+1}=p_{s+1} = 5$.
The identity (\ref{eq:BinomialFactorIdentity}) is easily proven from combinatorial considerations.
The proof is left to the reader.

\noindent{\bf The ``Sum Over a Simplex I'' Identity:}

\begin{eqnarray}
\sum\limits_{a < l_s < \dots < l_1 < b}
\prod\limits_{l=1}^{s} (l_{l})^{ n^{(l)}  }
=
\frac{ 
(b - a - 1)^{ s + N^{ s }   }
 }
{\prod\limits_{l=1}^{s} \left(l + (N^{ s }- N^{ s - l}) \right)}
\label{eq:SumOverSimplex}
\end{eqnarray}
where $N^{ j } := \sum\limits_{q=1}^j n^{(q)}$ for  $j=1,\dots,s$. 
The sum (\ref{eq:SumOverSimplex}) is done by proceeding from indices with large subscripts towards
those with small subscripts and at every time applying the identity (\ref{eq:SumIdentity}). 
In doing this we retain the highest order term only.
However, inclusion of lower terms is possible since they are of the same form as the highest order term.
It is only that the enumeration of all possible terms that emerge is cumbersome. This is left for future work.

\noindent{\bf The ``Sum Over a Simplex II'' Identity:}

\begin{eqnarray}
\lefteqn{\sum\limits_{0 \le n_0 \le  \dots \le  n_{p-1} \le  n_p} \prod\limits_{q=0}^{p-1} a_q^{n_q}}
\nonumber\\&&=
\sum\limits_{q=0}^p (-1)^q \frac{\prod\limits_{l=q}^{p-1} a_l^{n_p + p-l}}{\prod\limits_{l=0}^{q-1}(1-a_l\cdot \dots \cdot a_{q-1}) \prod\limits_{l=q}^{p-1}(1-a_q\cdot \dots \cdot a_{l})}
\label{eq:SumOverSimplexII}
\end{eqnarray}
The identity follows from a iterative application of the geometric sum formula.

\noindent{\bf The ``Integral Over a Simplex'' Identity:}

\begin{eqnarray}
\lefteqn{\int\limits_{0\le \xi_i \le \dots \le \xi_1 \le 1} d^i \xi
\vec{\xi}^{\vec{m}^{(1)}}
\cdot \prod\limits_{j=0}^{i-1}  \left(\xi_j - \xi_{j+1}\right)^{m^{(2)}_j}=}
\nonumber\\&&
\prod\limits_{j=0}^{i-1} m^{(2)}_j!
\cdot
\prod\limits_{j=1}^{i}
\frac{\left(j-1 + \sum\limits_{q=i-j+1}^i m_q^{(1)} + m_q^{(2)}  \right)}
{\left(j - m_{i-j}^{(1)} + \sum\limits_{q=i-j}^i m_q^{(1)} + m_q^{(2)} \right)}
\label{eq:IntegraloverSimplex}
\end{eqnarray}
where $\xi_{i+1} =0$ and $\xi_0 = 1$. 
The integral (\ref{eq:IntegraloverSimplex}) is computed by integrating in decreasing order of the subscript, i.e.\ starting from $\xi_i$ and ending at $\xi_1$, at each time substituting for $\xi_j = \xi_{j-1} t_j$ where $t_j \in [0,1]$ for $j=i,\dots,1$ and by using the identity :
\begin{equation}
\int\limits_0^1 dt t^n (1-t)^m = \frac{n! m!}{(n+m+1)!}
\end{equation}
for $n,m > -1$.

\noindent{\bf The ``Power of the Sum'' Identity:}

\begin{eqnarray}
\lefteqn{\left(\frac{1-z^{M+1}}{1-z} + z^j u\right)^N =} \nonumber\\
&& \sum\limits_{p=0}^M z^p \left( (\sum\limits_{q=0}^{N-1} 1_{q j \le p} u^q \alpha^{(p,j,N)}_q) + \delta_{N j,p} u^N\right)  + O(z^{M+1})\label{eq:PowerofSum}
\end{eqnarray}
where the coefficients read:
\begin{equation}
\alpha^{(p,j,N)}_q = C^N_q C^{p-q j + N-1 - q}_{N-1-q}
\end{equation}
for $q=0,\dots,N-1$.
The proof is in~\ref{AppB}.

\noindent{\bf The ``Differential'' Identity:}

\begin{equation}
\frac{d^m}{dy^m} \left(e^y-1\right)^q = \sum\limits_{s=1}^m 
\left(e^y-1\right)^{q-s} e^{s y} \left(\prod\limits_{j=0}^{s-1} (q-j)\right) a_s^{(m)}
\label{eq:DiffExp}
\end{equation}
where the coefficients satisfy recursion relations:
\begin{eqnarray}
a_s^{(m+1)} &=& s a_s^{(m)}1_{s\le m} + a_{s-1}^{(m)}1_{s\ge 2}
\nonumber\\
&=& \sum\limits_{\sum\limits_{l=0}^{s-1} n_l = m-s+1} \prod\limits_{l=0}^{s-1} (s-l)^{n_l}
\label{eq:Coefficients0} \\
& =&
 \sum\limits_{0\le \tilde{n}_0 \le \dots \le \tilde{n}_{s-2} \le m-s+1} (1)^{m-s+1} 
\prod\limits_{l=0}^{s-2} (\frac{s-l}{s-l-1})^{\tilde{n}_l}
\label{eq:Coefficients} \\
&=&
 \sum\limits_{q=0}^{s-1} (-1)^q 
\frac{\prod\limits_{l=q}^{s-2} (\frac{s-l}{s-l-1})^{m-l}}{\left(\prod\limits_{l=0}^{q-1} \frac{l-q}{s-q}\right)
\left(\prod\limits_{l=q}^{s-2} \frac{q-l-1}{s-l-1}\right)}
\label{eq:Coefficients1} \\
&=&
\frac{1}{(s-1)!}
\sum\limits_{q=0}^{s-1} (-1)^q
 C^{s-1}_q (s-q)^m
\label{eq:Coefficients2} \\
&=&
\{\{1\},\{1,1\},\{1,3,1\},\{1,7,6,1\},\{1,15,25,10,1\},\nonumber\\
&&\{1,31,90,65,15,1\}\,\dots\}
\label{eq:Coefficients3} 
\end{eqnarray}
with $a_s^{(1)} = 1$.
The equality in (\ref{eq:Coefficients0}) follows from iterating the recursion relation and recognizing a pattern in the consecutive
iterates.
 The equality in (\ref{eq:Coefficients}) follows from parameterizing the sum over the simplex and the equality in (\ref{eq:Coefficients1}) follows from
(\ref{eq:SumOverSimplexII}). Finally, the equality in (\ref{eq:Coefficients2}) results from simplifying 
expression (\ref{eq:SumOverSimplexII}).
In (\ref{eq:Coefficients3}) we give numerical values for the coefficients for $m=0,\dots,5$.
Note that since recursion relations such as (\ref{eq:Coefficients0}) appear in such vast fields as
anomalous diffusion processes and econometrics \& time series modelling, mathematical techniques
for solving these relations, presented here, are valuable for researchers from those fields.

\noindent{\bf The ``Polynomial'' Identity:}

\begin{equation}
\left(\alpha+1\right)_{n-1} := \prod\limits_{j=1}^n (\alpha+j) = 
\sum\limits_{p=0}^n \alpha^{n-p} \frac{n^{2p}}{2^p p!}
\label{eq:PolIdent}
\end{equation}
The identity (\ref{eq:PolIdent}) is valid in the limit $n\rightarrow \infty$ and it follows from
expanding the product into a sum and computing the coefficients at powers of $\alpha$
using identity (\ref{eq:SumOverSimplex}). 

\noindent{\bf An Auxiliary Lemma:}

\begin{eqnarray}
\!\!\!\!\!\! \lefteqn{S^{\alpha_1,\alpha_2}_{n;p_1,p_2} := \sum\limits_{n_1=p_1}^{n-p_2} (n-n_1)^{\alpha_1} n_1^{\alpha_2}}
\nonumber \\
\!\!\!\!\!\!\!\!\!\!\!\!\!\!\!\!\!\! &&=
(n-p_2)^{\alpha_2+1} n^{\alpha_1} \int\limits_0^1 d\xi \xi^{\alpha_2} (1-\frac{n-p_2}{n}\xi)^{\alpha_1} \nonumber \\
&& \mbox{}- p_1^{\alpha_2+1} n^{\alpha_1} \int\limits_0^1 d\xi \xi^{\alpha_2} (1-\frac{p_1}{n}\xi)^{\alpha_1}
\label{eq:AuxLemma} \\
\!\!\!\!\!\!\!\!\!\!\!\!\!\!\!\!\!\! &&= 
n^{\alpha_1+\alpha_2+1} \left(B(\frac{n-p_2}{n},\alpha_2+1,\alpha_1+1) - B(\frac{p_1}{n},\alpha_2+1,\alpha_1+1)\right)
\end{eqnarray}
Here $B$ is the incomplete Beta function.
The result (\ref{eq:AuxLemma}) follows from expanding the term in parentheses in a binomial expansion, doing the sum over
$n_1$ using the leading order term in (\ref{eq:SumIdentity}) and then re-summing the binomial expansion.

\noindent{\bf The Generalization of the ``Auxiliary Lemma'':}

\begin{eqnarray}
\!\!\!\!\!\!\!\!\!\!\!\!\!\!\!\!\!\!\!\!\!\!\!\! \lefteqn{S^{\alpha_,\dots,\alpha_N}_{n;p_1,\dots,p_N} :=
\sum\limits_{n_1+\dots+n_N=n} \prod\limits_{j=1}^N n_j^{\alpha_j} 1_{p_j \le n_j}}
\nonumber \\
\!\!\!\!\!\!\!\!\!\!\!\!\!\!\!\!\!\! &&=
n^{N-1+\sum\limits_{q=1}^N \alpha_q}
\int\limits_{[0,1]^{N-1}} d^{N-1}\theta
\left(\prod\limits_{j=1}^{N-2} \theta_j^{\sum\limits_{q=1}^{N-j} \alpha_q} (1-\theta_j)^{\alpha_{N-j+1}}\right)
\theta_{N-1}^{\alpha_2}(1-\theta_{N-1})^{\alpha_1}
\left(\prod\limits_{j=1}^{N-1} 
1_{\frac{p_{N-j}}{n \theta_{j-1}} \le \theta_j \le 1 - \frac{p_{N-j+1}}{n \theta_{j-1}}}
\right)
\label{eq:GenAuxLemma2}
\end{eqnarray}
The result (\ref{eq:GenAuxLemma2}) follows from an iterative application of (\ref{eq:AuxLemma})
and performing the manipulations that were used to derive (\ref{eq:AuxLemma}).
Note that when all $p_j = 0$ the integral on the right hand side factorizes and
equals the multivariate Beta function, which is 
$\prod\limits_{q=1}^N \frac{\alpha_q!}{(N-1+\sum\limits_{q=1}^N \alpha_q)}$.

\appsection{}\label{AppB} 

We prove the ``Power of the Sum'' identity (\ref{eq:PowerofSum}):
\begin{eqnarray}
\!\!\!\!\!\!\!\!\!\!\!\! \lefteqn{
\left(\frac{1-z^{M+1}}{1-z} + z^j u\right)^N = \left(\sum\limits_{p=1}^M z^p(1 + \delta_{p,j} u)\right)^N}
\nonumber \\
\!\!\!\!\!\!\!\!\!\!\!\! &&=
\left(\sum\limits_{p=0}^M z^p \sum\limits_{p_1=0}^p(\delta_{p_1,j}u+1)(\delta_{p-p_1,j}u+1)\right)^{N-1}
= \left(\cdots (\delta_{p,2j} u^2 + 2 \cdot1_{j\le p} u + p+1)\right)^{N-1}
\label{eq:PowerofSum1} \\
\!\!\!\!\!\!\!\!\!\!\!\! &&=
\left(\cdots \sum\limits_{p_1=0}^p (\delta_{p_1,j}u+1) (\delta_{p-p_1,2j} u^2 + 1_{j\le p-p_1} 2 u + C^{p-p_1+1}_1) \right)^{N-2}
\nonumber\\
\!\!\!\!\!\!\!\!\!\!\!\! &&=
\left(\cdots (\delta_{p,3j} u^3 + 3 \cdot1_{2j \le p} u^2 + 3 \cdot1_{j\le p} u C^{p-j+1}_1 + C^{p+2}_2)\right)^{N-3}
\label{eq:PowerofSum2} \\
\!\!\!\!\!\!\!\!\!\!\!\! &&=
\left(\cdots \sum\limits_{p_1=0}^p 
(\delta_{p_1,j}u+1)(\delta_{p-p_1,3j} u^3 + 3 \cdot1_{2j \le p-p_1}u^2 + 3\cdot 1_{j\le p-p_1} u C^{p-p_1-j+1}_1 + C^{p-p_1+2}_2)\right)^{N-4}
\nonumber \\
\!\!\!\!\!\!\!\!\!\!\!\! &&=
\left(\cdots (\delta_{p,4j} u^4 + 4 \cdot1_{3j\le p} u^3 + 6 \cdot1_{2j \le p} C^{p-2j+1}_1 u^2 + 4 \cdot1_{j\le p} C^{p-j+2}_2 + C^{p+3}_3) \right)^{N-4}
\end{eqnarray} 
Thus identity (\ref{eq:PowerofSum}) holds for $N=2,3,4$. Assume identity is valid for any value of $N$.
Then the coefficient at $z^p$ in $S^{N+1}$ reads:
\begin{eqnarray}
\!\!\!\!\!\!\!\!\!\!\!\! \lefteqn{\sum\limits_{p_1=0}^p (\delta_{p_1,j}u+1) \left(\sum\limits_{q=0}^{N-1} 1_{q j \le p-p_1} \alpha_q^{(p-p_1,j,N)} u^{q} + \delta_{N j,p- p_1} u^N\right) }
\label{eq:InductionProof} \\
\!\!\!\!\!\!\!\!\!\!\!\! & \!\!\!\!\!\!\!\!\!\!\!\! & \!\!\!\!\!\!\!\!\!\!\!\! =
\left(\sum\limits_{q=0}^{N-1} 1_{(q+1)j\le p} \alpha_q^{(p-j,j,N)} u^{q+1}\right) + \delta_{(N+1)j,p} u^{N+1} +
\left(\sum\limits_{q=0}^{N-1} \sum\limits_{p_1=0}^{p-qj} \alpha_q^{(p-p_1,j,N)} u^q\right) + 1_{N j \le p} u^N 
\nonumber \\
\!\!\!\!\!\!\!\!\!\!\!\! & \!\!\!\!\!\!\!\!\!\!\!\! & \!\!\!\!\!\!\!\!\!\!\!\! = \sum\limits_{q=0}^{N} (1_{q j\le p} C^N_{q-1} C^{p-q j+N-q}_{N-q} 1_{q\ge 1} + 1_{q< N}\sum\limits_{p_1=0}^{p-qj} C^N_q C^{p-p_1-qj+N-1-q}_{N-1-q} ) u^q +
1_{N j \le p} u^N + \delta_{(N+1)j,p} u^{N+1} 
\nonumber \\
\!\!\!\!\!\!\!\!\!\!\!\! & \!\!\!\!\!\!\!\!\!\!\!\! & \!\!\!\!\!\!\!\!\!\!\!\! = \sum\limits_{q=0}^{N} (1_{q j\le p} C^N_{q-1} C^{p-q j+N-q}_{N-q} 1_{q\ge 1} +  1_{q< N}C^N_q C^{p-qj+N-q}_{N-q} ) u^q +
1_{N j \le p} u^N + \delta_{(N+1)j,p} u^{N+1} 
\nonumber \\
\!\!\!\!\!\!\!\!\!\!\!\! & \!\!\!\!\!\!\!\!\!\!\!\! & \!\!\!\!\!\!\!\!\!\!\!\! = \sum\limits_{q=0}^{N} 1_{q j\le p} C^{N+1}_q C^{p-q j+N-q}_{N-q} u^q + \delta_{(N+1)j,p} u^{N+1}  \nonumber \\
\!\!\!\!\!\!\!\!\!\!\!\! & \!\!\!\!\!\!\!\!\!\!\!\! & \!\!\!\!\!\!\!\!\!\!\!\! =
\sum\limits_{q=0}^{N} 1_{q j \le p} \alpha_q^{(p,j,N+1)} u^{q} + \delta_{(N+1) j,p} u^{N+1} \quad\mbox{\bf q.e.d.} 
\label{eq:InductionProof2}
\end{eqnarray}

\appsection{}\label{AppC}

Here we derive a formula for the $p$-variate probability function 
of the occupation numbers, i.e.\ the likelihood to find
$n_{j_s}$ particles on energy level $j_s$ for $s=1,\dots,p$, respectively.
The calculations are a natural extension of calculations (\ref{eq:PDF0})-(\ref{eq:PDF3}), in the $p=1$ case
and thus we leave the explanations to the reader.
We take an ordered sequence $0\le j_1 < \dots < j_p$ and we write:
\begin{eqnarray}
\lefteqn{
C^{M+N-1}_{N-1} P\left(\bigcap\limits_{s=1}^p N_{j_s} = \tilde{n}_{j_s}\right) =
\sum\limits_{\sum\limits_{q=0}^M n_q=N} \frac{N!}{\prod\limits_{q=0}^M n_q!} \delta_{M,\sum\limits_{q=0}^M q n_q} 
\left(\prod\limits_{s=1}^p \delta_{n_{j_s},\tilde{n}_{j_s}}\right)}
\label{eq:PDFMultiVar0} \\
&&=
\left.\left.
\frac{1}{M!} \frac{d^M}{d z_1^M} 
\left(\prod\limits_{s=1}^p \frac{1}{\tilde{n}_{j_s}!} \frac{d^{\tilde{n}_{j_s}}}{dz_{1+s}^{\tilde{n}_{j_s}}}\right)
\left(\frac{1-z_1^{M+1}}{1-z_1} + (\sum\limits_{s=1}^p z_1^{j_s} (z_{1+s}-1))\right)^N
\right|_{z_1=0}
\right|_{z_{1+s}=0}
\label{eq:PDFMultiVar1} \\
&&=
\left(\prod\limits_{s=1}^p \frac{1}{\tilde{n}_{j_s}!} \frac{d^{\tilde{n}_{j_s}}}{dz_{1+s}^{\tilde{n}_{j_s}}}\right)
\left(
\sum\limits_{q=0}^{N-1} C^N_q 1_{J^{\vec{s}}_q \le M} C^{M- J^{\vec{s}}_q + N-1-q}_{N-1-q} 
\prod\limits_{l=1}^q (z_{1+s_l} - 1)
+
\delta_{M,J_N^{\vec{s}}} 
\prod\limits_{l=1}^N (z_{1+s_l} - 1)
\right)
\label{eq:PDFMultiVar2} \\
&&=
\left[
\sum\limits_{q=0}^{N-1}
1_{J^{\vec{s}}_q\le M}
\left(
\prod\limits_{l=1}^p C^{m_l^{(q)}}_{\tilde{n}_{j_l}} (-1)^{m_l^{(q)} - \tilde{n}_{j_l}} 1_{\tilde{n}_{j_l} \le m_l^{(q)}}
\right)
C^N_q C^{M - J^{\vec{s}}_q+N-1-q}_{N-1-q}
\right] 
\nonumber \\
&& \mbox{}+
\delta_{J^{\vec{s}}_q,M}
\left(
\prod\limits_{l=1}^p C^{m_l^{(N)}}_{\tilde{n}_{j_l}} (-1)^{m_l^{(N)} - \tilde{n}_{j_l}} 1_{\tilde{n}_{j_l} \le m_l^{(N)}}
\right)
\label{eq:PDFMultiVar3}
\end{eqnarray}
subject to $m_r^{(q)} = \sum\limits_{l=1}^q \delta_{r,s_l}$ 
and  $m_r^{(N)} = \sum\limits_{l=1}^N \delta_{r,s_l}$ for $r=1,\dots,p$.
Here \mbox{$J^{\vec{s}}_q := \sum\limits_{l=1}^q j_{s_l}$}.

\appsection{}\label{AppD}

Here we prove the large-$N$ limit of the multi-point probability function
$P\left(\bigcap\limits_{s=1}^p \left( N_{j_s}=\tilde{n}_{j_s}\right) \right)$ for the case of $j_s = s-1$.
Recall that $\left(j_l\right)_{l=1}^p$ is a strictly ascending sequence $0\le j_1 < \dots < j_p \le M$. 
Then, from (\ref{eq:CapJs}), $J^{\vec{s}}_q = (\sum\limits_{l=1}^q s_l ) - q$ and we have:
\begin{eqnarray}
\lefteqn{
\mathop{\mathop{\sum\limits_{s_1=1}^p\dots \sum\limits_{s_q=1}^p}_{\mbox{\tiny \# of $r$'s in $\left(s_l\right)_{l=1}^q$ = $m_r$ }}}_{\mbox{\tiny for $r=1,\dots,p$}}
 \frac{C^{M-J^{\vec{s}}+N-1-q}_{N-1-q}}{ C^{M+N-1}_{N-1}} }
\nonumber\\&&
\mathop{\rightarrow}_{N\rightarrow\infty}
\mathop{\mathop{\sum\limits_{s_1=1}^p\dots \sum\limits_{s_q=1}^p}_{\mbox{\tiny \# of $r$'s in $\left(s_l\right)_{l=1}^q$ = $m_r$ }}}_{\mbox{\tiny for $r=1,\dots,p$}}
 \frac{1}{T^q} \left(\frac{T}{T+1}\right)^{\sum\limits_{l=1}^q s_l}
\label{eq:LargeNLimMultVariate} \\
&&=\frac{1}{T^q} 
\frac{q!}{\prod\limits_{l=1}^p m_l!}
\left(\frac{T}{T+1}\right)^{\sum\limits_{l=1}^p l m_l}
\label{eq:LargeNLimMultVariate1} 
\end{eqnarray}
In (\ref{eq:LargeNLimMultVariate}) we used (\ref{eq:MthMomentLimit}) and (\ref{eq:MthMomentLimitResult}).
For simplicity we drop the superscript in the $m$ indices from now on.
Note that the sum runs over all integer grid-points of a $q$-dimensional hypercube of side length $p$
subject to the gridpoint having $m_r$ coordinates equal to $r=1,\dots,p$.
In (\ref{eq:LargeNLimMultVariate1}) we parametrized $\left(s_l\right)_{l=1}^q := \left(j\right)_{j=1,l=1}^{p_j^l,r}$ for $r =1,\dots,q$
subject to $\sum\limits_{l=1}^r p_\theta^{l} = m_\theta$ for $\theta =1,\dots,p$ with $p_\theta^{l} \ge 0$ and we summed the $p$ parameters.
Note that since the exponent $\sum\limits_{l=1}^q s_l=\sum\limits_{l=1}^p l m_l$ does not depend on the $p$  parameters the term in sum 
is multiplied by the cardinality of the set to be summed over and the cardinality in question equals the multinomial factor.
Thus, from (\ref{eq:PDFMultiVarMainText3}) and from (\ref{eq:LargeNLimMultVariate1}), we have:
\begin{eqnarray}
\lefteqn{P\left(\bigcap\limits_{s=1}^p N_{j_s}=\tilde{n}_{j_s}\right)=}
\label{eq:LargeNLimMultVariateFinal} \\
&&=
\sum\limits_{q=0}^N C^N_q \frac{(-1)^{q- \left|\vec{\tilde{n}}\right|}}{T^q}
\left(\frac{T}{T+1}\right)^{\sum\limits_{l=1}^p l \tilde{n}_{j_l}}
\frac{1}{\prod\limits_{l=1}^p \tilde{n}_{j_l}!}
\frac{q!}{(q-\left|\vec{\tilde{n}}\right|)!}
\left(
\frac{T}{T+1})^1 + \dots + (\frac{T}{T+1})^p
\right)^{q-\left|\vec{\tilde{n}}\right|}
\label{eq:LargeNLimMultVariateFinal1} \\
&&=
\frac{N!}{(N- \left|\vec{\tilde{n}}\right|)!(\prod\limits_{l=1}^p \tilde{n}_{j_l}!)}
\left((\frac{T}{T+1})^p\right)^{N-\left|\vec{\tilde{n}}\right|}
\prod\limits_{l=1}^p \left((\frac{T}{T+1})^l\frac{1}{T}\right)^{\tilde{n}_{j_l}}
\label{eq:LargeNLimMultVariateFinal2} \\
&&=
\frac{N!}{(N- \left|\vec{\tilde{n}}\right|)!(\prod\limits_{l=1}^p \tilde{n}_{j_l}!)} \prod\limits_{l=1}^{p+1} {\mathfrak p}_l^{\tilde{n}_{j_l}}
\label{eq:LargeNLimMultVariateFinal3} 
\end{eqnarray}
In (\ref{eq:LargeNLimMultVariateFinal1}) we defined $\left|\vec{\tilde{n}}\right| := \sum\limits_{l=1}^p \tilde{n}_{j_l}$ and 
in (\ref{eq:LargeNLimMultVariateFinal2}) we performed the sum over the $m^{(q)}_l$ parameters using the multinomial expansion formula.
In (\ref{eq:LargeNLimMultVariateFinal2}) we performed the sum over $q$ using the binomial expansion formula.
Thus, as seen in (\ref{eq:LargeNLimMultVariateFinal3}), the result is a multinomial distribution with likelihoods of individual trials given as:
\begin{equation}
{\mathfrak p}_l := 
\left\{
\begin{array}{ll}
(\frac{T}{T+1})^l\frac{1}{T} & \mbox{if $l=1,\dots,p$}\\
(\frac{T}{T+1})^p & \mbox{if $l=p$}
\end{array}
\right.
\label{eq:IndivProbs0}
\end{equation}
We check that the likelihoods of the individual trials sum up to unity. We have:
\begin{eqnarray}
\sum\limits_{l=1}^p {\mathfrak p}_l &=&
\frac{1}{T}
\left(\sum\limits_{l=1}^p (\frac{T}{T+1})^l\right)+
(\frac{T}{T+1})^p
\nonumber\\
&=&
1 - (\frac{T}{T+1})^p + (\frac{T}{T+1})^p = 1
\end{eqnarray}
as expected.

\appsection{}\label{AppE} 

Here we prove that the multi-point probability function (\ref{eq:PDFMultiVarMainText3})
is normalized to unity.
Firstly, for given $q$ we parametrize the sequence of $s$'s as follows
\mbox{$\left(s_l\right)_{l=1}^q := \left(\sum\limits_{\theta=1}^p \sum\limits_{\tilde{p}_\theta = 1}^{r_\theta} 
\delta_{l,k^\theta_{p_\theta}} \right)_{l=1}^q$}.
This means that we assume that the sequence of $s$'s contains
$r_i$ occurrences of the $i^{\mbox{th}}$ integer ($i=\{1, 2, \ldots, p \}$) at positions \mbox{$k^i_1,\dots,k^i_{\tilde{p}_1}$}.
Hence $J^{\vec{s}}_q = \sum\limits_{\theta=1}^p r_\theta j_\theta$ and $m^q_l = r_l$ for $l=1,\dots,p$ and
$m^q_l=0$ otherwise.
Inserting this into (\ref{eq:PDFMultiVarMainText3}) and summing over the occupation numbers we obtain:
\begin{eqnarray}
\lefteqn{
C^{M+N-1}_{N-1} \sum\limits_{\tilde{n}_{j_1},\dots,\tilde{n}_{j_p}} 
P\left(N_{j_1}=\tilde{n}_{j_1},\dots,N_{j_p}=\tilde{n}_{j_p}\right) = }
\nonumber \\
&&
\sum\limits_{q=0}^{N-1} 
\mathop{\sum\limits_{\sum\limits_{\theta=1}^p r_\theta = q}}_{r_1\ge 0,\dots,r_p\ge 0}
\left(
1_{\sum\limits_{\theta=1}^p r_\theta j_\theta \le M}
\frac{q!}{\prod\limits_{\theta=1}^p r_\theta!}
\prod\limits_{\theta=1}^p
\sum\limits_{\tilde{n}_{j_\theta}}
\left(
C^{r_\theta}_{\tilde{n}_{j_\theta}} (-1)^{r_\theta - \tilde{n}_{j_\theta}} 1_{\tilde{n}_{j_\theta} \le r_\theta}
\right)
\cdot
C^N_q C^{M-(\sum\limits_{\theta=1}^p r_\theta j_\theta)+N-1-q}_{N-1-q}
\right)
\nonumber\\
&& \mbox{}+
\mathop{\sum\limits_{\sum\limits_{\theta=1}^p r_\theta = N}}_{r_1\ge 0,\dots,r_p\ge 0}
\delta_{\sum\limits_{\theta=1}^p r_\theta j_\theta,M}
\frac{N!}{\prod\limits_{\theta=1}^p r_\theta!}
\prod\limits_{\theta=1}^p
\left(
\sum\limits_{\tilde{n}_{j_\theta}}
C^{r_\theta}_{\tilde{n}_{j_\theta}} (-1)^{r_\theta - \tilde{n}_{j_\theta}} 1_{\tilde{n}_{j_\theta} \le r_\theta}
\right) = \\
&&
\sum\limits_{q=0}^{N-1} 
\mathop{\sum\limits_{\sum\limits_{\theta=1}^p r_\theta = q}}_{r_1\ge 0,\dots,r_p\ge 0}
\left(
1_{\sum\limits_{\theta=1}^p r_\theta j_\theta \le M}
\frac{q!}{\prod\limits_{\theta=1}^p r_\theta!}
\prod\limits_{\theta=1}^p
(1-1)^{r_\theta}
\cdot
C^N_q C^{M-(\sum\limits_{\theta=1}^p r_\theta j_\theta)+N-1-q}_{N-1-q}
\right)
\nonumber\\
&&\mbox{}+
\mathop{\sum\limits_{\sum\limits_{\theta=1}^p r_\theta = N}}_{r_1\ge 0,\dots,r_p\ge 0}
\delta_{\sum\limits_{\theta=1}^p r_\theta j_\theta,M}
\frac{N!}{\prod\limits_{\theta=1}^p r_\theta!}
\prod\limits_{\theta=1}^p
(1-1)^{r_\theta} = \label{eq:MultiPointNormalization1}\\
&&
1_{0 \le M} C^{M+N-1}_{N-1} + \delta_{0,M} \delta_{0,N} = C^{M+N-1}_{N-1}
\label{eq:MultiPointNormalization2}
\end{eqnarray}
In (\ref{eq:MultiPointNormalization1}) we summed over the individual occupation numbers 
using the binomial expansion formula and the fact the term on the right hand side factorizes.
It is readily seen that out of the entire sum in (\ref{eq:MultiPointNormalization1}) 
only the term at $r_1=\dots=r_p=0$ along with $q=0$ and $N=0$ in the first and second expression respectively
is picked out which leads in a straightforward way to 
(\ref{eq:MultiPointNormalization1}). This finishes the proof.
 

\section{Figures}
\vspace{1cm}

\end{document}